%% file: main.tex
\documentclass{article}
\usepackage{float}

\usepackage[preprint]{neurips_2025}

\usepackage[utf8]{inputenc} 
\usepackage[T1]{fontenc}    
\usepackage{hyperref}       
\usepackage{url}            
\usepackage{booktabs}       
\usepackage{amsfonts}       
\usepackage{nicefrac}       
\usepackage{microtype}      
\usepackage{xcolor}         
\usepackage{color}
\usepackage{graphicx}
\usepackage{multirow}
\usepackage{algorithm}
\usepackage{algorithmic}
\usepackage{amsmath} 
\usepackage{bm}
\usepackage{pifont}

\title{CloneShield: A Framework for Universal Perturbation Against Zero-Shot Voice Cloning}

\author{
  Renyuan Li\textsuperscript{\dag}\And
  Zhibo Liang\textsuperscript{\dag}\And
  Haichuan Zhang\textsuperscript{}\And
  Tianyu Shi\textsuperscript{\dag}\And
  Zhiyuan Cheng\textsuperscript{}\And 
  Jia Shi\textsuperscript{}\And
  Carl Yang\textsuperscript{}\And
  Mingjie Tang\textsuperscript{\dag}
  \thanks{Authors marked with \dag\ are affiliated with Sichuan University. Contact: \texttt{tangrock@gmail.com}, \texttt{leonyuan73@gmail.com}.}
}

\begin{document}
\maketitle
\input{0_abstract}
\input{1_introduction}

\input{2_preliminaries}
\input{3_method}
\input{4_evaluation}

\input{5_ethics}
\input{6_conclusion}

{
\small
\bibliographystyle{plainnat}
\bibliography{main}
}


\newpage
\appendix

\section{Appendix}
\label{sec:appendix}

\subsection{Experimental Setup}
\paragraph{Defense setting}
In the initial experimental phase, adversarial perturbations were initialized as tensors, with their elements uniformly sampled from the interval $[-0.1, 0.1]$. Throughout the iterative defense process, the $L_{\infty}$ norm of the generated perturbations was constrained to a maximum of $0.15$. For defenses conducted on three distinct models, the number of iterations was consistently set to $60$.

In the second stage of the defense, we set different initial coefficients for the loss. For YourTTS, the initial coefficients for the reference loss and output loss were set to [7/200] and [1/600], respectively. For XTTSv2, the initial coefficients for the reference loss and output loss were set to [1/100] and [3/1000], respectively. For IndexTTS, we set the coefficients for the reference loss and output loss to [3/25] and [13/100], respectively.we also set different saving strategies for different models. For YourTTS, we directly saved the noise after 300 iterations. For XTTSv2, we set a threshold for the reference loss at 1.65 and then saved the result that achieved the minimum output loss among those iterations where the reference loss was below this threshold. For IndexTTS, we saved the result with the minimum output loss from the final thirty iterations. For these second-stage defense experiments on all three models, the number of iterations was set to 60. Additionally, a uniform initial learning rate of 0.001 was applied across all three models. The learning rate was managed by an identical StepLR (Step Learning Rate) scheduler for these experiments, configured to reduce the learning rate by a factor of 0.7  every 30 steps .
For the computation of the loss on the Mel spectrogram representation of the input audio, we utilized three distinct configurations for the transformation. These configurations were designed to capture features at different resolutions. Subsequent to generating the Mel spectrograms with each configuration---which inherently represent spectral power---these were converted into a decibel (dB) scaled representation. This conversion employed a  value of 80.0 to effectively manage the dynamic range, yielding the features upon which the loss was computed. The configurations are detailed as follows:
\paragraph{Large Mel Spectrogram Configuration}This transform employed a Fast Fourier Transform (FFT) window size of 2048 points, with a hop length of 512 samples between successive frames. The number of Mel filterbanks was set to 80.
\paragraph{Small Mel Spectrogram Configuration}For this setting, while maintaining the same number of 80 Mel filterbanks, the FFT window size was reduced to 512 points, and the hop length was set to 128 samples. 
\paragraph{Medium Mel Spectrogram Configuration}The medium configuration used 80 Mel filterbanks. The FFT window size was further reduced to 1024 points, with a corresponding hop length of 256 samples. These varying parameters for FFT window size and hop length allow for the analysis of the audio signal at different time-frequency resolutions, which can be beneficial for robust loss calculation.
\subsection{Comprehensive experimental results}
\paragraph{Ablation Study for Varying Perturbation Magnitudes}
In the initial phase of our experiments, we conducted an ablation study to assess the impact of varying perturbation magnitudes. The corresponding results are presented below. These findings led to the selection of 0.15 as the final perturbation magnitude for subsequent experiments.Based on an empirical evaluation of various perturbation strengths, as detailed in Table \ref{tab:Various Perturbation Settings}, we selected a perturbation strength of $\epsilon=0.15$ for our primary experiments. This value was determined to strike an optimal balance between maintaining the quality of the perturbed input audio and achieving the desired level of content and speaker identity obfuscation in the subsequent Text-to-Speech (TTS) output. Specifically, at the $\epsilon=0.15$ strength, the perturbed input audio (designated as Type 'in' in Table\ref{tab:Various Perturbation Settings}) consistently demonstrated higher perceptual quality and fidelity compared to inputs generated with other tested strengths. For instance, across different batch configurations, inputs at $\epsilon=0.15$ generally exhibited higher PESQ scores (approx. 4.17--4.26), higher SRS values (approx. 0.954--0.957) indicating better preservation of the original speaker's characteristics in the perturbed input, and lower MCD values (approx. 3.1--3.5) suggesting less spectral distortion relative to a clean signal baseline. These metrics collectively indicate minimal perceptible difference introduced at the input stage at this perturbation level.
Concurrently, the TTS outputs generated from these $\epsilon=0.15$ perturbed inputs (designated as Type 'out') achieved a highly effective degree of obfuscation. This was evidenced by substantially degraded output quality metrics (e.g. PESQ $\approx$ 1.07--1.09) and, critically, significantly lower SRS scores (approx. 0.269--0.289). These low output SRS values signify a strong deviation from the original speaker identity, thereby fulfilling our content protection objectives. This level of output modification was comparable to, or in some aspects such as output SRS, more effective than that achieved with other perturbation strengths, especially when considering the superior quality and lower perceptibility of the perturbation at the $\epsilon=0.15$ input stage. Thus, $\epsilon=0.15$ was chosen for its ability to deliver effective output transformation while minimizing alterations to the input audio.
Further examination of the TTS output characteristics (designated as Type 'out' in Table \ref{tab:Various Perturbation Settings}) when the input audio is perturbed at the $\epsilon=0.15$ strength reinforces this selection as optimal. The primary goal for the output is effective content and speaker identity obfuscation. At $\epsilon=0.15$, metrics assessing the synthesized speech demonstrate a profound transformation. For instance, PESQ scores for the output are consistently low (approx. 1.07--1.09 across batch configurations at $\epsilon=0.15$), and STOI values are substantially negative (approx. -16 to -14), both indicating significant degradation in perceptual quality and intelligibility, which is a desired outcome for the defense.
More critically, the speaker recognition similarity (SRS) for these outputs at $\epsilon=0.15$ is markedly low (ranging from approximately 0.269 to 0.289). This is comparable to, and in several configurations, lower (i.e., better for obfuscation) than the output SRS values achieved at other perturbation strengths (e.g., average SRS for $\epsilon=0.1$ 'out' is $\approx 0.30$; for $\epsilon=0.3$ 'out' is $\approx 0.29$; for $\epsilon=0.5$ 'out' is $\approx 0.29$). This demonstrates a highly effective scrambling of the original speaker's vocal characteristics. While all tested perturbation strengths lead to substantially altered outputs, the $\epsilon=0.15$ level achieves this strong defensive posture for the output with a compellingly low degree of alteration to the input audio itself. This combination solidifies $\epsilon=0.15$ as the perturbation strength that provides the most advantageous trade-off for robust content protection.

\paragraph{Ablation Study for Second Optimization Stage}
Furthermore, we conducted an ablation study to investigate the impact of the second optimization stage on the characteristics of the adversarial perturbations. As detailed in Table \ref{tab:before_optimization_comparison_filled} and Table \ref{tab:after_optimization_comparison_filled}, the results indicate that the inclusion of this second stage significantly enhances both the imperceptibility of the perturbations and their defense effectiveness against the target TTS model.

\begin{table*}[htbp]
\centering
\caption{Evaluation Metrics under Various Perturbation Settings, $\epsilon$ is the bound of perturbation} 
\label{tab:Various Perturbation Settings} 
\resizebox{\textwidth}{!}{%
\begin{tabular}{@{}ll|ccccccc@{}} 
\toprule\toprule
\multicolumn{1}{c}{\textbf{Batch}} & \multicolumn{1}{c|}{\textbf{Type}} & MCD & PESQ & STOI & SDR & LSD & SNR & SRS \\ 
\midrule

\multicolumn{9}{c}{\textbf{$\epsilon=0.1$}} \\ 
\midrule
\multirow{2}{*}{1 per batch} & in  & 3.851 & 2.974 & 0.964 & 10.925 & 2.426 & 7.619 & 0.897 \\
                             & out & 14.722 & 1.110 & 0.235 & -14.086 & 4.794 & -4.773 & 0.336 \\
\cmidrule(lr){1-9} 
\multirow{2}{*}{3 per batch} & in  & 3.999 & 3.156 & 0.969 & 11.934 & 2.360 & 7.789 & 0.902 \\
                             & out & 15.786 & 1.105 & 0.134 & -16.171 & 5.490 & -4.563 & 0.279 \\
\cmidrule(lr){1-9}
\multirow{2}{*}{5 per batch} & in  & 3.640 & 3.087 & 0.969 & 11.667 & 2.392 & 8.180 & 0.899 \\
                             & out & 15.329 & 1.079 & 0.166 & -14.994 & 5.559 & -4.863 & 0.284 \\
\midrule\midrule

\multicolumn{9}{c}{\textbf{$\epsilon=0.15$}} \\
\midrule
\multirow{2}{*}{1 per batch} & in  & 3.324 & 4.202 & 0.987 & 18.376 & 2.382 & 10.449 & 0.957 \\
                             & out & 15.978 & 1.091 & 0.153 & -14.957 & 4.832 & -5.290 & 0.289 \\
\cmidrule(lr){1-9}
\multirow{2}{*}{3 per batch} & in  & 3.502 & 4.259 & 0.990 & 19.392 & 2.323 & 9.924 & 0.956 \\
                             & out & 15.890 & 1.066 & 0.127 & -15.578 & 5.380 & -4.760 & 0.269 \\
\cmidrule(lr){1-9}
\multirow{2}{*}{5 per batch} & in  & 3.143 & 4.174 & 0.989 & 19.071 & 2.349 & 10.650 & 0.954 \\
                             & out & 15.499 & 1.072 & 0.172 & -14.386 & 5.504 & -5.004 & 0.275 \\
\midrule\midrule

\multicolumn{9}{c}{\textbf{$\epsilon=0.3$}} \\
\midrule
\multirow{2}{*}{1 per batch} & in  & 3.853 & 2.941 & 0.963 & 10.877 & 2.426 & 7.614 & 0.892 \\
                             & out & 14.993 & 1.102 & 0.218 & -13.756 & 4.870 & -4.853 & 0.339 \\
\cmidrule(lr){1-9}
\multirow{2}{*}{3 per batch} & in  & 3.995 & 3.158 & 0.969 & 11.949 & 2.359 & 7.811 & 0.903 \\
                             & out & 15.765 & 1.076 & 0.123 & -15.233 & 5.702 & -4.627 & 0.267 \\
\cmidrule(lr){1-9}
\multirow{2}{*}{5 per batch} & in  & 3.666 & 3.053 & 0.968 & 11.647 & 2.393 & 8.163 & 0.900 \\
                             & out & 15.346 & 1.083 & 0.173 & -14.858 & 5.735 & -4.954 & 0.270 \\
\midrule\midrule

\multicolumn{9}{c}{\textbf{$\epsilon=0.5$}} \\
\midrule
\multirow{2}{*}{1 per batch} & in  & 3.881 & 2.947 & 0.962 & 10.896 & 2.425 & 7.604 & 0.899 \\
                             & out & 15.903 & 1.117 & 0.147 & -14.280 & 5.348 & -4.822 & 0.335 \\
\cmidrule(lr){1-9}
\multirow{2}{*}{3 per batch} & in  & 4.007 & 3.155 & 0.969 & 11.922 & 2.360 & 7.788 & 0.901 \\
                             & out & 15.633 & 1.076 & 0.152 & -15.235 & 5.447 & -4.599 & 0.272 \\
\cmidrule(lr){1-9}
\multirow{2}{*}{5 per batch} & in  & 3.652 & 3.054 & 0.969 & 11.660 & 2.393 & 8.179 & 0.899 \\
                             & out & 15.514 & 1.094 & 0.158 & -15.161 & 5.555 & -4.974 & 0.271 \\
\bottomrule
\end{tabular}%
} 
\end{table*}

\begin{table*}[htbp]
\centering
\caption{Comparison of Evaluation Metrics Before Stage2 under Different Datasets, Word Error Rate(WER)}
\label{tab:before_optimization_comparison_filled}
\resizebox{\textwidth}{!}{%
\begin{tabular}{@{}lll|ccccccccc@{}}
\toprule\toprule
\multicolumn{1}{c}{\textbf{Dataset}} & \multicolumn{1}{c}{\textbf{Batch}} & \multicolumn{1}{c|}{\textbf{Type}} & WER & CER & MCD & PESQ & STOI & SDR & LSD & SNR & SRS \\
\midrule
\multicolumn{12}{c}{\textbf{Before Stage2}} \\
\midrule
\multirow{6}{*}{\textit{LibriSpeech}} & \multirow{2}{*}{1 per batch} & in  & 0.226 & 0.073 & 1.602 & 2.045 & 0.962 & 18.573 & 1.132 & 16.534 & 0.726 \\
                                  &                              & out & 0.250 & 0.109 & 5.686 & 1.071 & 0.207 & -16.109 & 3.844 & -0.013 & 0.347 \\
                                  \cmidrule(lr){2-12}
                                  & \multirow{2}{*}{3 per batch} & in  & 0.669 & 0.608 & 2.621 & 2.289 & 0.969 & 19.094 & 1.087 & 12.031 & 0.805 \\
                                  &                              & out & 0.175 & 0.090 & 5.149 & 1.100 & 0.171 & -15.958 & 2.519 & -0.057 & 0.501 \\
                                  \cmidrule(lr){2-12}
                                  & \multirow{2}{*}{5 per batch} & in  & 0.436 & 0.328 & 3.262 & 2.344 & 0.972 & 19.620 & 1.099 & 10.624 & 0.807 \\
                                  &                              & out & 0.118 & 0.057 & 4.560 & 1.108 & 0.193 & -14.881 & 2.920 & -0.094 & 0.569 \\
\midrule
\multirow{6}{*}{\textit{CommonVoice}} & \multirow{2}{*}{1 per batch} & in  & 0.393 & 0.212 & 1.948 & 1.643 & 0.949 & 18.366 & 5.683 & 16.512 & 0.624 \\
                                  &                              & out & 0.033 & 0.019 & 16.692 & 1.083 & 0.189 & -15.802 & 2.927 & -0.024 & 0.439 \\
                                  \cmidrule(lr){2-12}
                                  & \multirow{2}{*}{3 per batch} & in  & 0.315 & 0.160 & 2.913 & 1.680 & 0.948 & 18.445 & 5.296 & 11.044 & 0.664 \\
                                  &                              & out & 0.053 & 0.022 & 16.695 & 1.122 & 0.167 & -16.489 & 2.609 & -0.093 & 0.591 \\
                                  \cmidrule(lr){2-12}
                                  & \multirow{2}{*}{5 per batch} & in  & 0.375 & 0.201 & 3.344 & 1.709 & 0.955 & 18.594 & 5.299 & 9.176 & 0.686 \\
                                  &                              & out & 0.012 & 0.005 & 16.587 & 1.112 & 0.184 & -15.742 & 2.884 & -0.157 & 0.642 \\
\midrule
\multirow{6}{*}{\textit{LibriTTS}} & \multirow{2}{*}{1 per batch} & in  & 0.179 & 0.056 & 1.971 & 2.088 & 0.970 & 18.523 & 2.535 & 16.351 & 0.720 \\
                                  &                              & out & 0.028 & 0.013 & 15.598 & 1.050 & 0.194 & -15.117 & 3.020 & -0.008 & 0.357 \\
                                  \cmidrule(lr){2-12}
                                  & \multirow{2}{*}{3 per batch} & in  & 0.144 & 0.043 & 2.827 & 2.137 & 0.973 & 18.982 & 2.480 & 12.433 & 0.765 \\
                                  &                              & out & 0.073 & 0.042 & 15.036 & 1.062 & 0.164 & -15.917 & 2.694 & -0.059 & 0.523 \\
                                  \cmidrule(lr){2-12}
                                  & \multirow{2}{*}{5 per batch} & in  & 0.152 & 0.049 & 3.632 & 2.215 & 0.975 & 19.002 & 2.601 & 9.538 & 0.784 \\
                                  &                              & out & 0.011 & 0.008 & 14.636 & 1.070 & 0.183 & -15.107 & 2.458 & -0.100 & 0.566 \\
\midrule
\multirow{6}{*}{\textit{VCTK}} & \multirow{2}{*}{1 per batch} & in  & 0.119 & 0.040 & 1.504 & 1.715 & 0.901 & 18.219 & 2.483 & 16.503 & 0.630 \\
                                  &                              & out & 0.020 & 0.004 & 14.402 & 1.087 & 0.232 & -18.191 & 2.959 & -0.026 & 0.421 \\
                                  \cmidrule(lr){2-12}
                                  & \multirow{2}{*}{3 per batch} & in  & 0.107 & 0.040 & 2.762 & 1.804 & 0.904 & 18.376 & 2.565 & 9.792 & 0.688 \\
                                  &                              & out & 0.027 & 0.008 & 14.296 & 1.085 & 0.218 & -17.379 & 2.668 & -0.140 & 0.553 \\
                                  \cmidrule(lr){2-12}
                                  & \multirow{2}{*}{5 per batch} & in  & 0.115 & 0.045 & 3.246 & 1.851 & 0.918 & 18.415 & 2.619 & 7.948 & 0.714 \\
                                  &                              & out & 0.044 & 0.012 & 14.095 & 1.103 & 0.219 & -16.979 & 2.332 & -0.161 & 0.585 \\
\midrule
\multirow{6}{*}{\textit{LJSpeech}} & \multirow{2}{*}{1 per batch} & in  & 0.101 & 0.043 & 2.257 & 1.664 & 0.963 & 17.991 & 2.685 & 15.155 & 0.665 \\
                                  &                              & out & 0.006 & 0.002 & 16.996 & 1.072 & 0.202 & -15.378 & 3.376 & -0.011 & 0.414 \\
                                  \cmidrule(lr){2-12}
                                  & \multirow{2}{*}{3 per batch} & in  & 0.110 & 0.054 & 2.596 & 1.812 & 0.970 & 18.282 & 2.702 & 13.219 & 0.762 \\
                                  &                              & out & 0.041 & 0.040 & 15.961 & 1.067 & 0.212 & -15.923 & 2.783 & -0.088 & 0.604 \\
                                  \cmidrule(lr){2-12}
                                  & \multirow{2}{*}{5 per batch} & in  & 0.123 & 0.044 & 2.734 & 1.872 & 0.972 & 18.482 & 2.713 & 12.487 & 0.787 \\
                                  &                              & out & 0.000 & 0.000 & 15.330 & 1.087 & 0.209 & -15.162 & 2.505 & -0.149 & 0.662 \\
\bottomrule
\end{tabular}%
} 
\end{table*}

\begin{table*}[htbp]
\centering
\caption{Comparison of Evaluation Metrics After Stage2 under Different Datasets}
\label{tab:after_optimization_comparison_filled} 
\resizebox{\textwidth}{!}{%
\begin{tabular}{@{}lll|ccccccccc@{}}
\toprule\toprule
\multicolumn{1}{c}{\textbf{Dataset}} & \multicolumn{1}{c}{\textbf{Batch }} & \multicolumn{1}{c|}{\textbf{Type}} & WER & CER & MCD & PESQ & STOI & SDR & LSD & SNR & SRS \\
\midrule
\multicolumn{12}{c}{\textbf{After Stage2}} \\
\midrule
\multirow{6}{*}{\textit{LibriSpeech}} & \multirow{2}{*}{1 per batch} & in  & 0.220 & 0.069 & 3.177 & 3.624 & 0.968 & 17.026 & 0.933 & 10.037 & 0.909 \\
                                  &                              & out & 1.108 & 0.850 & 18.480 & 1.123 & 0.145 & -21.977 & 8.587 & -0.789 & 0.074 \\
                                  \cmidrule(lr){2-12}
                                  & \multirow{2}{*}{3 per batch} & in  & 0.211 & 0.063 & 2.948 & 3.887 & 0.975 & 18.013 & 0.933 & 10.854 & 0.923 \\
                                  &                              & out & 1.093 & 0.803 & 20.145 & 1.078 & 0.169 & -20.310 & 8.098 & -2.344 & 0.076 \\
                                  \cmidrule(lr){2-12}
                                  & \multirow{2}{*}{5 per batch} & in  & 0.214 & 0.067 & 3.163 & 3.912 & 0.976 & 18.205 & 0.945 & 10.205 & 0.927 \\
                                  &                              & out & 1.144 & 0.823 & 20.033 & 1.081 & 0.166 & -20.150 & 8.034 & -2.483 & 0.078 \\
\midrule
\multirow{6}{*}{\textit{CommonVoice}} & \multirow{2}{*}{1 per batch} & in  & 0.376 & 0.210 & 3.692 & 3.430 & 0.967 & 18.348 & 5.011 & 8.751 & 0.880 \\
                                  &                              & out & 0.947 & 1.043 & 17.932 & 1.132 & 0.127 & -18.444 & 5.700 & -1.808 & 0.050 \\
                                  \cmidrule(lr){2-12}
                                  & \multirow{2}{*}{3 per batch} & in  & 0.321 & 0.174 & 3.228 & 3.644 & 0.972 & 18.638 & 4.577 & 10.105 & 0.906 \\
                                  &                              & out & 1.150 & 0.971 & 20.129 & 1.106 & 0.126 & -16.798 & 6.235 & -2.618 & 0.055 \\
                                  \cmidrule(lr){2-12}
                                  & \multirow{2}{*}{5 per batch} & in  & 0.346 & 0.175 & 3.382 & 3.711 & 0.974 & 18.811 & 4.501 & 9.679 & 0.914 \\
                                  &                              & out & 1.010 & 0.860 & 20.390 & 1.129 & 0.119 & -16.534 & 6.244 & -2.794 & 0.056 \\
\midrule
\multirow{6}{*}{\textit{LibriTTS}} & \multirow{2}{*}{1 per batch} & in  & 0.177 & 0.058 & 3.864 & 3.654 & 0.970 & 17.077 & 2.328 & 9.506 & 0.905 \\
                                  &                              & out & 1.008 & 0.861 & 19.488 & 1.210 & 0.119 & -22.413 & 8.875 & -1.139 & 0.036 \\
                                  \cmidrule(lr){2-12}
                                  & \multirow{2}{*}{3 per batch} & in  & 0.141 & 0.042 & 3.591 & 3.820 & 0.976 & 17.821 & 2.321 & 9.878 & 0.917 \\
                                  &                              & out & 1.056 & 0.821 & 20.760 & 1.215 & 0.136 & -19.885 & 8.249 & -2.205 & 0.039 \\
                                  \cmidrule(lr){2-12}
                                  & \multirow{2}{*}{5 per batch} & in  & 0.152 & 0.047 & 3.458 & 3.821 & 0.977 & 17.872 & 2.373 & 10.287 & 0.924 \\
                                  &                              & out & 1.034 & 0.813 & 21.213 & 1.136 & 0.128 & -20.386 & 8.210 & -2.388 & 0.041 \\
\midrule
\multirow{6}{*}{\textit{VCTK}} & \multirow{2}{*}{1 per batch} & in  & 0.105 & 0.034 & 2.586 & 3.530 & 0.937 & 17.187 & 2.325 & 10.540 & 0.863 \\
                                  &                              & out & 1.288 & 0.945 & 17.497 & 1.188 & 0.122 & -22.881 & 8.093 & -1.855 & 0.055 \\
                                  \cmidrule(lr){2-12}
                                  & \multirow{2}{*}{3 per batch} & in  & 0.104 & 0.037 & 2.380 & 4.179 & 0.965 & 18.561 & 2.314 & 10.535 & 0.946 \\
                                  &                              & out & 0.289 & 0.132 & 13.986 & 1.115 & 0.228 & -14.565 & 4.726 & -4.664 & 0.483 \\
                                  \cmidrule(lr){2-12}
                                  & \multirow{2}{*}{5 per batch} & in  & 0.116 & 0.047 & 2.721 & 3.902 & 0.953 & 17.701 & 2.323 & 9.703 & 0.913 \\
                                  &                              & out & 1.154 & 0.851 & 19.429 & 1.138 & 0.182 & -17.119 & 7.662 & -4.117 & 0.050 \\
\midrule
\multirow{6}{*}{\textit{LJSpeech}} & \multirow{2}{*}{1 per batch} & in  & 0.106 & 0.045 & 3.596 & 3.292 & 0.971 & 17.779 & 2.629 & 10.006 & 0.893 \\
                                  &                              & out & 0.988 & 0.648 & 18.912 & 1.123 & 0.110 & -18.361 & 8.274 & -0.323 & 0.011 \\
                                  \cmidrule(lr){2-12}
                                  & \multirow{2}{*}{3 per batch} & in  & 0.115 & 0.054 & 3.693 & 3.589 & 0.979 & 18.554 & 2.665 & 9.490 & 0.924 \\
                                  &                              & out & 0.899 & 0.558 & 20.455 & 1.105 & 0.159 & -18.511 & 7.416 & -1.999 & 0.040 \\
                                  \cmidrule(lr){2-12}
                                  & \multirow{2}{*}{5 per batch} & in  & 0.119 & 0.043 & 3.574 & 3.702 & 0.981 & 18.957 & 2.661 & 9.823 & 0.928 \\
                                  &                              & out & 0.924 & 0.558 & 20.949 & 1.112 & 0.166 & -17.952 & 7.063 & -2.712 & 0.045 \\
\bottomrule
\end{tabular}%
} 
\end{table*}

\subsection{Subjective audio evaluation}
\subsubsection{Evaluation of Perceptual Quality using Mean Opinion Score (MOS)}
To assess the perceptual quality of audio signals in our study, we employ the Mean Opinion Score (MOS). MOS is a widely recognized and standardized subjective measure that quantifies the human-perceived quality of speech and audio\citep{ITU-T:P800:2016}. It relies on human listeners to rate the quality of test samples. These ratings are typically provided on a five-point Absolute Category Rating (ACR) scale. The final MOS for a given audio sample or condition is then calculated as the arithmetic mean of the scores assigned by the panel of listeners. If $S_k$ represents the score given by the $k$-th listener for a particular audio sample, and there are $N$ listeners in total, the MOS is calculated as:
$$ \text{MOS} = \frac{1}{N} \sum_{k=1}^{N} S_k $$
This metric provides a direct and intuitive measure of subjective quality. In this work, we comprehensively evaluate perceptual quality by employing both traditional subjective MOS listening tests with human participants and objective quality estimations derived from the advanced DNN-based audio predictor, UTMOSv2\citep{baba2024t05}. The specific experimental settings for each of these evaluation methodologies are detailed in the subsequent sections.

\paragraph{Participants and Listening Environment}
Sixty participants (N=60), primarily undergraduate and graduate students from the Computer Science Department at our university, were recruited for this study. All participants reported normal hearing. Prior to the main experiment, each participant listened to a short English audio segment from a blog to ensure adequate comprehension of the language used in the test stimuli, as our source audio content was in English. The listening tests were conducted in a professional recording studio on the university campus, providing a controlled acoustic environment with minimal ambient noise (<30 dBA). Participants used high-quality, noise-canceling headphones Sony WH-1000XM5 for audio playback, with volume levels calibrated to a comfortable and consistent listening level across all sessions.

\paragraph{Stimuli and Experimental Design}

The original English speech segments used as the basis for our evaluation were randomly selected from our dataset. For the listening tests, we then focused on evaluating only the following specific versions derived from each of these selected original audio segments:
\begin{enumerate}
    \item \textbf{Original Audio (O):} The unprocessed, clean speech signal (referred to as the "original dataset" sample).
    \item \textbf{Audio Perturbed by Our Method (P-Ours):} Audio signals with imperceptible perturbations added by our proposed adversarial perturbation method (referred to as the "input of our method").
    \item \textbf{TTS Synthesized Speech from Original Audio (TTS-O):} Speech synthesized by the target Text-to-Speech (TTS) model using the Original Audio as input (referred to as the "model's original output").
    \item \textbf{TTS Synthesized Speech from Our Perturbed Audio (TTS-P-Ours):} Speech synthesized by the target Text-to-Speech (TTS) model using the audio perturbed by our method (P-Ours) as input (referred to as the "model's output using our method's input").
\end{enumerate}
Participants were presented with these stimuli in a randomized order to mitigate context and order effects. For each participant, 10 distinct sets, each containing these four types of audio samples derived from an original segment, were evaluated to ensure a comprehensive assessment.

\paragraph{Rating Procedure and Data Analysis}

Participants were tasked with evaluating the overall audio quality of each presented stimulus. Prior to commencing the evaluation, all participants received detailed instructions on the rating procedure and the 5-point Absolute Category Rating (ACR) Mean Opinion Score (MOS) scale, which is a classic international standard ITU-T P.800. The scale was explicitly defined as follows:
\begin{itemize}
    \item \textbf{5: Excellent}
    \item \textbf{4: Good}
    \item \textbf{3: Fair}
    \item \textbf{2: Poor}
    \item \textbf{1: Bad}
\end{itemize}
Listeners provided their rating for each audio sample immediately after its presentation. Specifically, this procedure was applied to assess the following two categories of stimuli:
\begin{enumerate}
    \item The original unprocessed audio (\textbf{O}) and the audio perturbed by our proposed method (\textbf{P-Ours}).
    \item The TTS synthesized speech outputs generated from the original audio (\textbf{TTS-O}) and from the audio perturbed by our method (\textbf{TTS-P-Ours}).
\end{enumerate}
For each of these four conditions (O, P-Ours, TTS-O, TTS-P-Ours), Mean Opinion Scores (MOS) were calculated by averaging the scores from all participants. We also computed the 95\% confidence intervals for these mean scores to indicate the precision of the subjective ratings. The 95\% confidence interval (CI) for a mean MOS score is calculated using the formula:
\[ \text{CI} = \left[ \overline{\text{MOS}} - 1.96 \times \frac{\sigma}{\sqrt{N}}, \overline{\text{MOS}} + 1.96 \times \frac{\sigma}{\sqrt{N}} \right] \]
where $\overline{\text{MOS}}$ is the calculated sample mean MOS, $\sigma$ is the standard deviation of the scores from the participants, and $N$ is the number of participants (in this case, N=60).The results are in Table \ref{tab:mos_results_ci}

\begin{table}[htbp] 
\centering
\caption{Mean Opinion Scores (MOS) with 95\% Confidence Intervals (N=60)}
\label{tab:mos_results_ci}
\begin{tabular}{@{}lcc@{}}
\toprule
\textbf{Evaluation Target} & \textbf{MOS Mean} & \textbf{95\% CI Range} \\
\midrule
O                       & 3.10 & (2.94, 3.26) \\
P-Ours       & 2.80 & (2.62, 2.98) \\
TTS-P-Ours     & 0.20 & (0.09, 0.31) \\ 
TTS-O                      & 3.05 & (2.88, 3.22) \\
\bottomrule
\end{tabular}
\end{table}

\subsubsection{Objective MOS Prediction with UTMOSv2}

For objective evaluation of speech synthesis quality, we employed UTMOSv2, 
an advanced Mean Opinion Score (MOS) prediction system specifically designed 
for assessing the naturalness of high-quality synthetic speech\citep{baba2024t05}. 
UTMOSv2  
adopts a sophisticated approach by combining features from diverse sources. 
It leverages a pretrained Self-Supervised Learning (SSL) model (wav2vec 2.0) 
to extract speech representations and, distinctively, incorporates a 
pretrained deep image classifier (EfficientNetV2) to capture subtle 
differences in speech spectrograms. 
These SSL-based and spectrogram-based features are then integrated through 
a feature fusion mechanism and a multi-stage fine-tuning strategy to 
optimize MOS prediction accuracy.

The performance of UTMOSv2 has been rigorously evaluated in competitive 
benchmarks. Notably, in the VoiceMOS Challenge 2024 Track 1, 
which focused on predicting MOS for high-quality synthetic speech, 
particularly in "zoomed-in" listening test scenarios, the UTMOSv2 system 
demonstrated state-of-the-art results. 
It achieved first place in 7 out of the 16 official evaluation metrics 
and secured second place in the remaining 9 metrics\citep{baba2024t05}. 
This top-tier performance was reported to be significantly better than 
systems ranked third or lower, underscoring its effectiveness (Abstract and Section~1 of\citep{baba2024t05}. 
Further ablation studies within the original work confirmed that the 
fusion of spectrogram and SSL features was crucial for improving 
correlation-based evaluation metrics with human judgments (Section~4.4.2 of\citep{baba2024t05}). 
This strong empirical performance highlights UTMOSv2's capability to 
provide reliable and accurate objective MOS predictions, making it a 
valuable tool for automated assessment in demanding speech synthesis 
evaluation contexts. The results calculated using UTMOSv2 are in Table \ref{tab:utmos_detailed_scores_30samples_avg_textdata}.

\begin{table*}[htbp]
\centering
\caption{Detailed UTMOS Scores for Various Systems and Conditions (30 Samples and Average)}
\label{tab:utmos_detailed_scores_30samples_avg_textdata} 
\resizebox{\textwidth}{!}{
\begin{tabular}{@{}lcccccccccc@{}} 
\toprule
& \multicolumn{2}{c}{\textbf{Benign}} & \multicolumn{2}{c}{\textbf{Voicebox}} & \multicolumn{2}{c}{\textbf{AudioSeal}} & \multicolumn{2}{c}{\textbf{Timbre Watermarking}} & \multicolumn{2}{c}{\textbf{Ours}} \\
\cmidrule(lr){2-3} \cmidrule(lr){4-5} \cmidrule(lr){6-7} \cmidrule(lr){8-9} \cmidrule(lr){10-11}
\textbf{Sample \#} & \textbf{In} & \textbf{Out} & \textbf{In} & \textbf{Out} & \textbf{In} & \textbf{Out} & \textbf{In} & \textbf{Out} & \textbf{In} & \textbf{Out} \\
\midrule
1  & 3.1309 & 2.7070 & 2.4590 & 1.9697 & 2.5527 & 2.3281 & 3.0996 & 2.2813 & 2.0078 & 0.5103 \\
2  & 3.2324 & 2.0313 & 3.1855 & 1.9219 & 3.4648 & 2.3477 & 3.5723 & 2.0527 & 2.0898 & 0.4543 \\
3  & 2.2188 & 2.6836 & 1.8242 & 2.0020 & 2.3906 & 2.8516 & 2.8301 & 2.0098 & 1.7217 & 0.5298 \\
4  & 1.6387 & 1.6963 & 1.4473 & 3.2559 & 2.2793 & 1.9219 & 1.7021 & 1.9795 & 1.7002 & 0.4333 \\
5  & 2.5410 & 3.0078 & 2.3027 & 2.1426 & 3.0840 & 2.6094 & 2.8711 & 2.6406 & 1.7227 & 0.5425 \\
6  & 2.7012 & 2.4355 & 1.7695 & 2.7012 & 2.1641 & 2.4727 & 1.9678 & 2.2129 & 1.9844 & 0.0186 \\
7  & 3.3184 & 2.7285 & 3.3008 & 2.4238 & 3.3633 & 2.8125 & 3.0625 & 2.7461 & 3.1289 & 0.4353 \\
8  & 2.3496 & 2.8047 & 2.1504 & 2.3066 & 1.9785 & 2.1621 & 2.1895 & 2.3223 & 1.9766 & 0.0666 \\
9  & 2.9453 & 3.4473 & 3.1680 & 2.6602 & 3.1602 & 2.7324 & 3.1426 & 2.9668 & 2.6387 & 0.4346 \\
10 & 1.3076 & 2.6348 & 1.7002 & 2.4766 & 1.2803 & 2.2520 & 1.9932 & 2.0430 & 1.6904 & 0.5586 \\
11 & 2.2910 & 2.7930 & 2.1211 & 2.8281 & 2.1738 & 2.5039 & 2.2715 & 2.4707 & 1.8047 & 0.1163 \\
12 & 3.1191 & 3.3438 & 2.6445 & 2.0664 & 2.7461 & 2.7910 & 2.8203 & 2.8691 & 2.2227 & 0.1260 \\
13 & 1.8994 & 2.3145 & 1.7402 & 2.5547 & 2.4043 & 2.4141 & 2.3262 & 2.7109 & 1.6416 & 0.5601 \\
14 & 2.6211 & 2.9785 & 2.4648 & 2.3750 & 2.6992 & 2.6992 & 2.2734 & 3.2969 & 1.9883 & 0.2944 \\
15 & 2.8027 & 1.9189 & 2.2656 & 2.4336 & 2.3984 & 2.5742 & 2.6055 & 2.5234 & 1.9395 & 0.1857 \\
16 & 2.2891 & 2.3242 & 2.5645 & 2.2656 & 2.8184 & 2.6816 & 2.4668 & 2.2578 & 2.0781 & 0.0252 \\
17 & 1.7793 & 2.7090 & 1.4453 & 2.9824 & 2.0605 & 2.5313 & 1.4375 & 2.8672 & 1.5117 & 0.5352 \\
18 & 1.9355 & 3.0195 & 1.4268 & 2.0117 & 2.3203 & 2.8086 & 2.1387 & 2.7988 & 1.8770 & 0.5444 \\
19 & 3.0898 & 2.2109 & 3.0918 & 2.5664 & 2.9785 & 2.9688 & 2.9316 & 1.8281 & 2.4316 & 0.2620 \\
20 & 2.9863 & 2.9297 & 2.9551 & 2.8359 & 2.9180 & 2.8438 & 1.8447 & 1.7783 & 2.5742 & 0.1594 \\
21 & 2.5020 & 2.3887 & 2.5410 & 2.3184 & 2.6113 & 2.7148 & 1.9766 & 2.2656 & 2.1348 & -0.1055 \\
22 & 2.8633 & 2.0938 & 3.0801 & 1.9854 & 2.8672 & 2.2363 & 2.8887 & 2.3340 & 2.0176 & 0.0286 \\
23 & 2.2656 & 2.0742 & 2.5156 & 2.4199 & 1.9658 & 1.9033 & 1.8506 & 1.6104 & 2.1660 & 0.0836 \\
24 & 2.2266 & 2.0313 & 2.1113 & 2.6387 & 2.2773 & 2.4082 & 1.8447 & 2.4531 & 1.9961 & 0.2161 \\
25 & 2.1387 & 2.2148 & 1.4609 & 3.0488 & 1.7041 & 1.5215 & 1.8418 & 1.8643 & 1.5156 & 0.3398 \\
26 & 2.1719 & 2.3340 & 1.9941 & 1.8926 & 2.2930 & 2.6719 & 1.8604 & 2.4688 & 1.9600 & 0.1970 \\
27 & 2.4004 & 1.5137 & 1.8906 & 2.8027 & 2.5234 & 1.6807 & 2.4180 & 2.0371 & 1.5742 & 0.1722 \\
28 & 2.0605 & 1.7979 & 1.5840 & 2.3262 & 1.6934 & 2.1016 & 1.6025 & 1.6914 & 1.4053 & 0.3416 \\
29 & 1.8965 & 2.2148 & 1.4492 & 3.0273 & 2.3125 & 2.1230 & 1.9365 & 2.0156 & 1.8457 & 0.8618 \\
30 & 2.0020 & 2.4727 & 1.7998 & 2.9434 & 2.0234 & 1.8652 & 1.9922 & 2.0273 & 1.8994 & 0.5972 \\
\midrule
\textbf{Average} & \textbf{2.4242} & \textbf{2.4727} & \textbf{2.2151} & \textbf{2.4728} & \textbf{2.4502} & \textbf{2.4178} & \textbf{2.3263} & \textbf{2.3141} & \textbf{1.9748} & \textbf{0.3174} \\
\bottomrule
\end{tabular}%
} 
\end{table*}

\subsection{Audio objective index details}
\textbf{Objective Evaluation Metrics}
To comprehensively assess the impact of our proposed method on audio signals, we employed a suite of objective metrics targeting various aspects of speech: content intelligibility, audio quality/distortion, and speaker identity.

\textbf{Word Error Rate (WER) and Character Error Rate (CER)}
WER and CER are standard metrics for evaluating the accuracy of Automatic Speech Recognition (ASR) systems\citep{Povey_ASRU2011}. They quantify the discrepancies between the reference transcript and the ASR hypothesis; lower scores indicate better intelligibility and content preservation. Both metrics are derived from the Levenshtein distance, computed as the sum of substitutions (S), deletions (D), and insertions (I) required to transform the hypothesis into the reference, normalized by the total number of words ($N_W$) or characters ($N_C$) in the reference transcript:
\[ \text{WER} = \frac{S_W + D_W + I_W}{N_W} \]
\[ \text{CER} = \frac{S_C + D_C + I_C}{N_C} \]
In our experiments, ASR was performed using the OpenAI Whisper 'base' model. WER and CER were then computed using the \texttt{jiwer} library by comparing the ASR outputs (of original and processed audios) against the reference transcripts.

\textbf{Mel-Cepstral Distortion (MCD)}
 Mel-Cepstral Distortion (MCD) is a widely adopted objective metric for quantifying the spectral dissimilarity between two audio signals, particularly in evaluating the timbral quality of synthesized or converted speech against a reference. It operates by comparing the Mel Frequency Cepstral Coefficients (MFCCs) extracted from the two signals. A lower MCD value signifies a higher degree of spectral similarity, indicating that the processed speech is closer in timbral characteristics to the reference speech\citep{Chen_2022_CVPR}.

The calculation involves extracting MFCC vectors, $\mathcal{C} = \{c_1, c_2, \ldots, c_T\}$ from the processed audio and $\mathcal{C}' = \{c'_1, c'_2, \ldots, c'_T\}$ from the reference audio, where $T$ is the number of frames (assuming both signals are aligned or processed to have the same number of frames). For each frame $t$, the Euclidean distance $d(c_t, c'_t)$ between the corresponding $K$-dimensional MFCC vectors (typically $K=13$, excluding the 0th coefficient) is computed.  The overall MCD is then the average of these per-frame distances:
\[ \text{MCD}(\mathcal{C},\mathcal{C}') = \frac{1}{T}\sum_{t=1}^{T}d(c_{t},c_{t}') = \frac{1}{T}\sum_{t=1}^{T}\sqrt{\sum_{k=1}^{K}(c_{t,k}-c_{t,k}')^{2}} \]
 where $c_{t,k}$ and $c'_{t,k}$ represent the $k$-th coefficient of the MFCC vectors at frame $t$ for the processed and reference speech, respectively. In our experiments, this metric was calculated between two audio using the \texttt{pymcd} library (``plain'' mode), which aligns with this formulation when input lengths are matched. This metric was calculated between the original reference audio and the processed test audio using the \texttt{pymcd} library.

\textbf{Perceptual Evaluation of Speech Quality (PESQ)}
PESQ is an objective metric standardized in ITU-T Recommendation P.862 for predicting the subjective listening quality of speech\citep{cernak2005evaluation}. It is widely used for evaluating speech codecs and transmission networks, with higher scores indicating better perceptual quality. PESQ compares a reference signal to a degraded signal. This process involves several stages, including temporal alignment, level equalization, and a psychoacoustic model to transform the signals into an internal representation that mimics human auditory perception. The differences in this domain are then mapped to a quality score, typically ranging from -0.5 to 4.5.
Our PESQ scores were computed between the original reference audio and the processed test audio using the \texttt{pesq} library, operating in wideband ('wb') mode with audio signals at 16\,kHz.

\textbf{Short-Time Objective Intelligibility (STOI)}
STOI is an objective measure designed to predict the intelligibility of speech degraded by additive noise or distortions. Higher STOI scores, ranging from 0 to 1, correlate with better speech intelligibility. STOI operates by comparing the short-time temporal envelope correlations between the clean reference and the degraded speech signals across several one-third octave frequency bands.\citep{taal2011algorithm}The average of these correlations forms the STOI score.
We calculated STOI between the original reference audio and the processed test audio using the \texttt{pystoi} library. Signals were processed at 16\,kHz, and the standard STOI formulation (\texttt{extended=False}) was used.

\textbf{Signal-to-Distortion Ratio (SDR)}
SDR quantifies the level of a target signal relative to unwanted distortions, which can include noise, interference, and artifacts introduced by processing. It is a common metric in audio source separation and speech enhancement, where higher SDR values (in dB) indicate less distortion. SDR is computed by decomposing the processed signal into components corresponding to the true target signal, interference, noise, and artifacts. It is the ratio of the power of the target signal component to the power of the sum of the other undesired components.
SDR was calculated using the \texttt{bss\_eval\_sources} function from the \texttt{mir\_eval} library\citep{Raffel2014mir_eval}, comparing the original reference audio against the processed test audio, with signals at 16\,kHz.

\textbf{Log-Spectral Distance (LSD)}
LSD, also known as log-spectral distortion, measures the average dissimilarity between the log-magnitude spectra of two audio signals\citep{Gray1976DistanceMeasures}; lower values indicate greater spectral similarity and less distortion. LSD is computed by taking the root mean square error (RMSE) between the log-magnitude spectra of the reference and test signals, typically averaged over frames and frequency bins. Given the short-time Fourier transforms $S_1(t,f)$ and $S_2(t,f)$ for two signals:
\[ \text{LSD} = \text{mean}_t \left( \sqrt{\text{mean}_f \left( (\log|S_1(t,f)| - \log|S_2(t,f)|)^2 \right)} \right) \]
The values are often reported in dB. LSD was computed between the original reference audio and the processed test audio using STFTs generated by \texttt{librosa} (FFT window: 2048, hop length: 512).

\textbf{Signal-to-Noise Ratio (SNR)}
SNR is a fundamental measure of signal strength relative to background noise\citep{Oppenheim1999DSP}, with higher SNR values (in dB) indicating a cleaner signal with less noise contamination. In our context, the reference signal is considered the clean signal, and the difference between the reference and the degraded signal is treated as noise. SNR is then calculated as:
\[ \text{SNR} = 10 \log_{10} \left( \frac{P_{\text{signal}}}{P_{\text{noise}}} \right) = 10 \log_{10} \left( \frac{\sum |s_{\text{ref}}(n)|^2}{\sum |s_{\text{ref}}(n) - s_{\text{deg}}(n)|^2} \right) \]
SNR was calculated between the original reference audio and the processed test audio, with signals processed at 16\,kHz.

\textbf{Speaker Reconition Similarity (SRS)}
SRS is a metric designed to quantify the speaker identity similarity between two audio samples. This is particularly relevant for assessing whether an defense alters speaker characteristics or if TTS outputs retain the target speaker's voice; higher scores indicate greater speaker similarity. We employed a pre-trained speaker embedding model (\texttt{pyannote/embedding})\citep{Plaquet23}\citep{Bredin23} to extract embeddings from the original reference audio and the processed test audio (using \texttt{pyannote.audio} with the ``whole'' window setting). The SRS is then computed as the cosine similarity between these two speaker embedding vectors:
\[ \text{SRS} = \frac{\mathbf{e}_1 \cdot \mathbf{e}_2}{\|\mathbf{e}_1\| \|\mathbf{e}_2\|} \]
where $\mathbf{e}_1$ and $\mathbf{e}_2$ are the speaker embeddings. Scores range from -1 to 1, with values closer to 1 indicating higher confidence in identical or very similar speaker identities.

\subsection{Additional Spectrograms}
This section contains additional spectral visualizations to supplement the main paper. The figures include standard Spectrograms,showing the log-magnitude STFT on a linear frequency scale (dB), and Log-Mel Spectrograms, showing the log-power on a mel frequency scale.
To aid in visual differentiation in the presented figures, the standard Spectrograms are rendered with a lighter, more varied color palette, while the Log-Mel Spectrograms utilize a darker, sequential color palette. These figures provide further examples of the audio features analyzed.

\begin{figure}[htp] 
  \centering 
  \includegraphics[width=\linewidth]{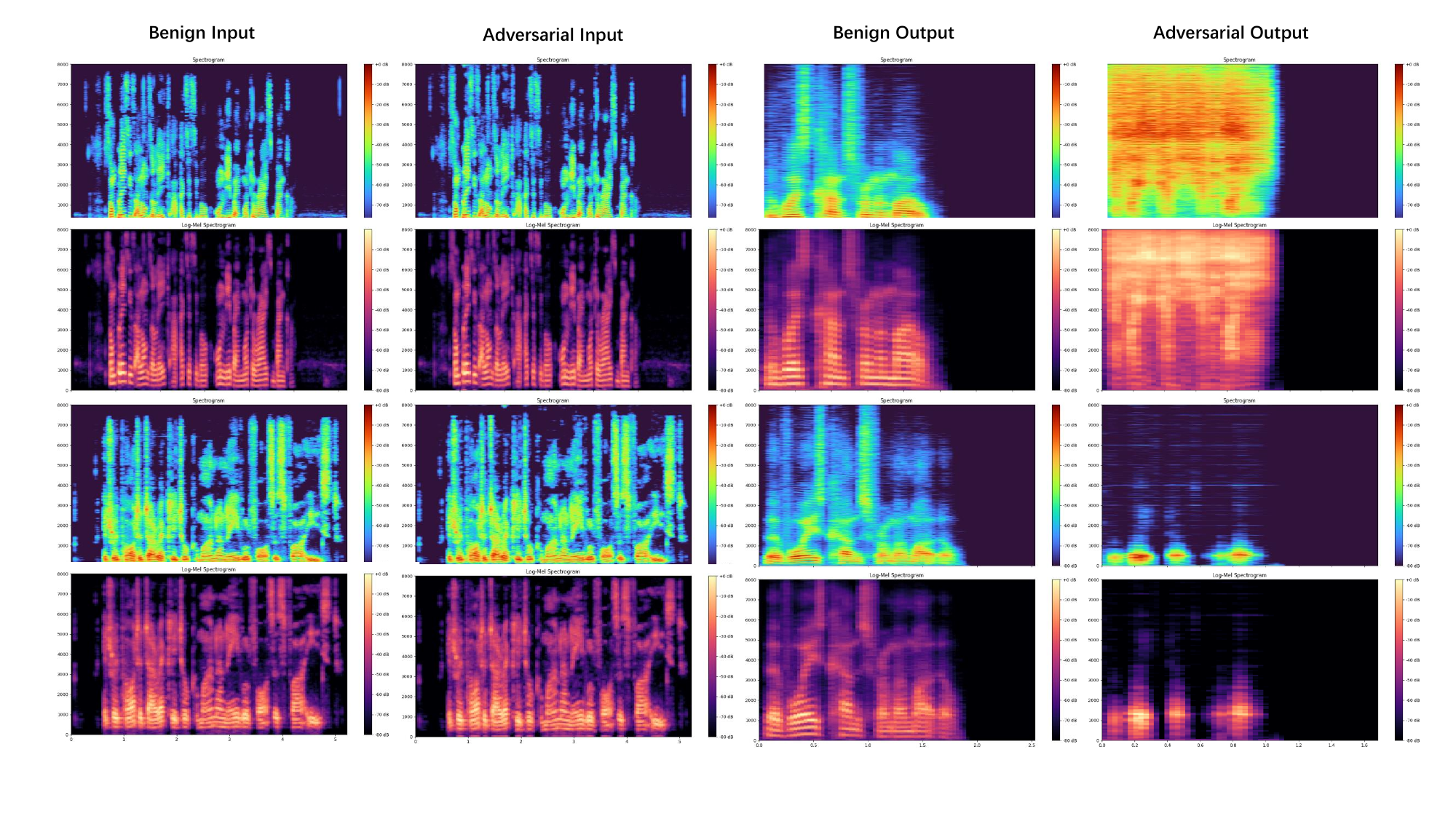}
   \includegraphics[width=\linewidth]{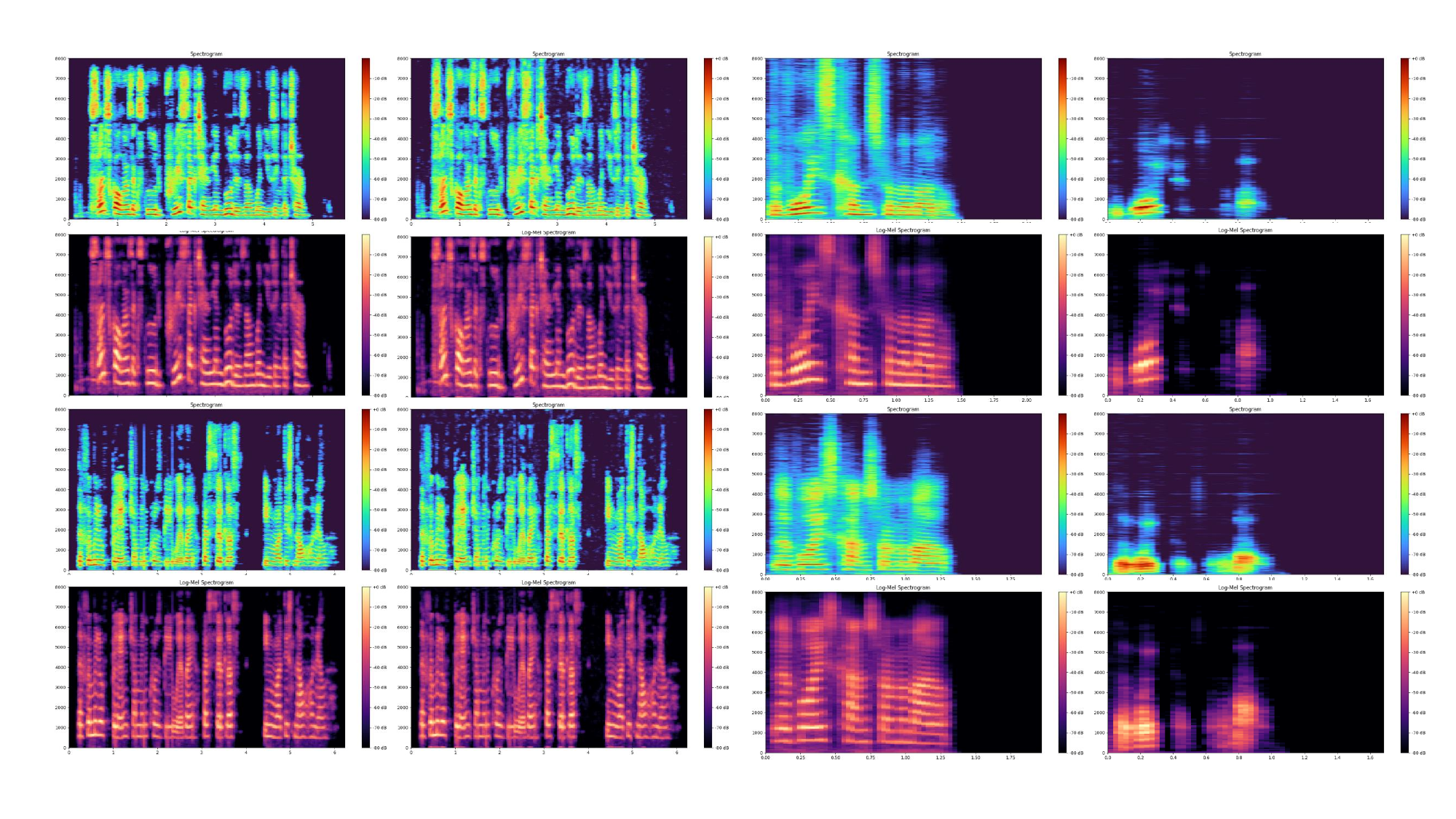}
  \caption{Additional Spectrograms}
  \label{fig:spectrogram_example}
\end{figure}

\end{document}

%% file: 0_abstract.tex
\begin{abstract}

Recent breakthroughs in text-to-speech (TTS) voice cloning have raised serious privacy concerns, allowing highly accurate vocal identity replication from just a few seconds of reference audio, while retaining the speaker’s vocal authenticity. 
In this paper, we introduce \textbf{CloneShield}, a universal time-domain adversarial perturbation framework specifically designed to defend against zero-shot voice cloning. Our method provides protection that is robust across speakers and utterances, without requiring any prior knowledge of the synthesized text.
We formulate perturbation generation as a multi-objective optimization problem, and propose Multi-Gradient Descent Algorithm (MGDA) to ensure the robust protection across diverse utterances. To preserve natural auditory perception for users, we decompose the adversarial perturbation via Mel-spectrogram representations and fine-tune it for each sample. This design ensures imperceptibility while maintaining strong degradation effects on zero-shot cloned outputs.
Experiments on three state-of-the-art zero-shot TTS systems, five benchmark datasets and evaluations from 60 human listeners demonstrate that our method preserves near-original audio quality in protected inputs (PESQ = 3.90, SRS = 0.93) while substantially degrading both speaker similarity and speech quality in cloned samples (PESQ = 1.07, SRS = 0.08). 

\end{abstract}

%% file: 1_introduction.tex
\section{Introduction}
\label{sec:intro}
Text-to-Speech (TTS) models are capable of producing natural and expressive speech by capturing essential acoustic characteristics~\citep{jeong2021diff,lajszczak2024base,li2024styletts}. With the emergence of large-scale pre-trained generative models~\citep{harshvardhan2020comprehensive,zeni2023mattergen,spens2024generative,rombach2022ldm}, TTS systems now exhibit remarkable performance and have been widely deployed in various real-world scenarios, including customer service automation~\citep{wang2023voice,roslan2023rise}, digital content creation, and accessibility technologies~\citep{sun2024intelligent,janokar2023text}. 
A particularly powerful capability of modern TTS systems is zero-shot voice cloning~\citep{dai2022cloning,guennec2023voice}, which enables accurate replication of a speaker's vocal characteristics and speaking style from just seconds of reference audio. Such advancement significantly lowers the barrier to vocal identity replication~\citep{casanova2022yourtts,casanova2024xtts,deng2025indextts,tan2024naturalspeech}.
However, the rise of voice synthesis technologies poses serious privacy and security threats. With only a few seconds of public audio, attackers can impersonate individuals for fraud or misinformation. Naturally, protecting against unauthorized voice cloning is more urgent than ever~\citep{wang2025one}.

Despite recent advances in text-to-speech (TTS) and voice synthesis, existing voice protection techniques fall short in defending against zero-shot voice cloning. Although text-to-speech (TTS) and voice synthesis technologies have made significant strides, current voice protection techniques remain inadequate, especially against zero-shot voice cloning attacks. Prior approaches such as audio watermarking~\citep{liu2023detecting, san2024proactive}, adversarial perturbations targeting ASR systems~\citep{le2024voicebox}, and defenses designed for voice conversion (VC)~\citep{li2023voice, huang2021defending}, either operate post-synthesis, require paired training data, or depend on prior knowledge of the input content. These assumptions break down in open-domain cloning scenarios, where attackers can clone arbitrary voices without access to transcripts or aligned corpora. Consequently, there remains a pressing need for proactive and content-agnostic protection mechanisms tailored to the zero-shot setting.

In this paper, we propose a proactive defense method that disrupts zero-shot voice cloning during the inference stage of TTS models. As shown in Figure~\ref{fig:framework}, the core idea is inspired by adversarial audio techniques: By injecting carefully crafted perturbations into a group of benign speech clips, we significantly impair the quality of cloned speech while preserving the naturalness and intelligibility of the original audio for human listeners.
However, crafting such perturbations is inherently challenging. At first, even when targeting a single utterance from a specific speaker, crafting an effective perturbation to resist voice cloning is non-trivial. It requires carefully injecting imperceptible noise that degrades the cloned output while preserving perceptual quality.
In addition, in realistic deployment scenarios, protecting each utterance individually is inefficient and computationally expensive. Therefore, we take a step further: we aim to defend a batch of utterances from different speakers using a shared perturbation strategy. While final perturbations may still be fine-tuned per sample, they are all derived from a common base update, which introduces a much more difficult problem—ensuring consistent protection effectiveness across diverse speaker identities, speaking styles, and acoustic conditions.
Second, the perturbation must remain imperceptible to human listeners, such that the protected audio is acoustically indistinguishable from the original in terms of naturalness, intelligibility, and overall quality.
Naturally, introducing adversarial noise—no matter how subtle—can still degrade perceptual quality, resulting in audible artifacts or distortions. To mitigate this, we introduce a lightweight post-processing step that smooths the perturbed audio and restores fidelity without compromising its protective effect. This step helps preserve a high-quality listening experience while maintaining robustness against cloning attacks.

To address these challenges, our method introduces two key innovations. First, we employ a multi-objective optimization strategy to derive a universal perturbation across multiple samples. Second, we refine the perturbation in the perceptual-frequency domain to strike a balance between attack strength and perceptual transparency. In summary, we make the following contributions.




\begin{itemize}
    \item We propose a framework that generates a single, imperceptible time-domain perturbation to protect a group of utterances against zero-shot voice cloning attacks. The perturbation generalizes across multiple utterances and does not require access to target texts or cloned output.
    \item We formulate the generation of universal perturbations as a multi-objective optimization problem and adopt the Multi-Gradient Descent Algorithm (MGDA) to jointly optimize the perturbation across multiple audio samples, ensuring effective speaker protection while maintaining audio fidelity.
    \item To further preserve the perceptual quality of protected audios, we introduce a perceptual-frequency domain refinement strategy that decomposes the perturbation for each protected utterance, enabling adaptive adjustments that preserve human listening quality while maintaining defense success.
    \item We conducted extensive experiments on five benchmark datasets under three TTS models. The results show that the protected audios remain perceptually close to the original speaker (speaker recognition similarity (SRS): $0.93$), while the cloned outputs exhibit significantly disrupted speaker identity (SRS: $0.08$), and our defense success rate (DSR) is $100\%$.
    \item We discuss the ethical considerations and broader societal impacts of our proposed defense, highlighting its role in promoting responsible voice AI development and privacy protection.
    
\end{itemize}

%% file: 2_preliminaries.tex
\begin{figure*}[htbp]
  \centering
   \includegraphics[width=1.0\linewidth]{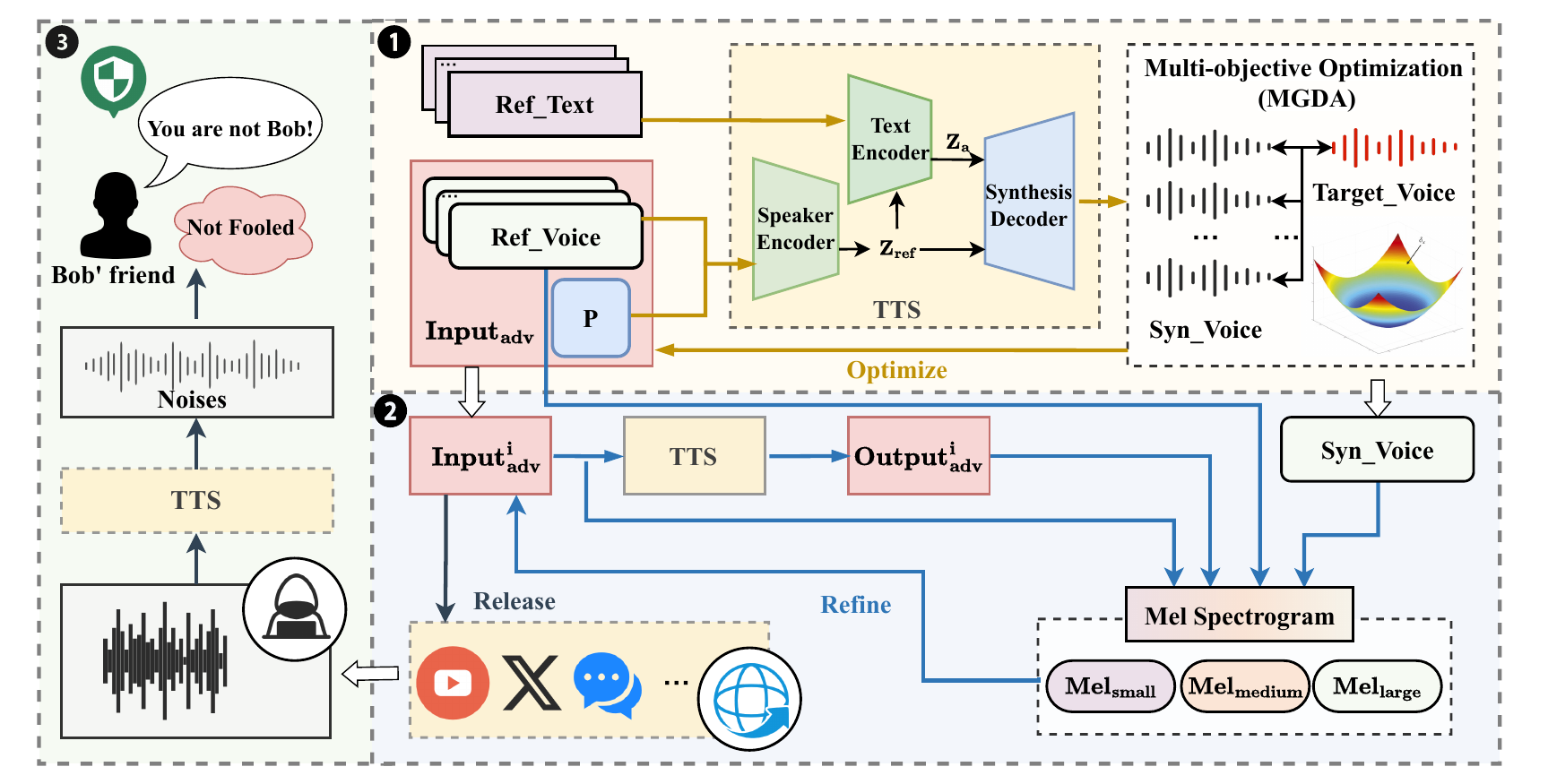}
   \caption{
Overview of our \textsc{CloneShield} framework. We inject imperceptible perturbations to disrupt the unauthorized voice replication. The system consists of \ding{182}Universal protective perturbation generation via multi-objective optimization. \ding{183}Perceptual-frequency domain refinement/fine-tune using mel-spectrogram decomposition. \ding{184}Real-world deployment scenarios showcasing how the perturbation thwarts unauthorized voice replication.
}
   \label{fig:framework}
\end{figure*} \vspace{-0.4em}

\section{Related Works}\vspace{-0.6em}
\subsection{Text-to-Speech Models}
\label{subsec:tts_models}
\vspace{-0.6em}
The emergence of large language models (LLMs) has accelerated the evolution of text-to-speech systems from traditional pipeline architectures to end-to-end neural approaches, and more recently to LLM-powered zero-shot cloning systems~\citep{naveed2023comprehensive,chang2024survey,popov2021grad,touvron2023llama,touvron2023llama2,popov2021grad,mehta2024matcha}. Early neural models such as Tacotron 2~\citep{elias2021parallel} and FastSpeech~\citep{ren2020fastspeech} introduced sequence-to-sequence frameworks with attention and duration prediction, significantly improving speech quality and synthesis speed.
Later, VITS~\citep{kim2021conditional} proposed a fully end-to-end probabilistic model that integrates text encoding, duration modeling, and waveform generation into a single variational framework, enabling faster inference and more expressive speech. VITS became a foundational architecture for many subsequent TTS systems.
Building on VITS, YourTTS~\citep{casanova2022yourtts} introduced multilingual and zero-shot speaker adaptation capabilities. It demonstrated that high-quality voice cloning can be achieved using just a few seconds of reference audio, without requiring speaker-specific training. YourTTS has since inspired numerous follow-up works.
XTTS~\citep{casanova2024xtts} extended YourTTS with an autoregressive decoder, improving the naturalness and controllability of the generated speech. It supports multilingual voice cloning and is widely used in both academia and commercial platforms.
Most recently, IndexTTS~\citep{deng2025indextts} achieved state-of-the-art performance by incorporating large language model features, FSQ-based quantization, and a conformer-based encoder. It outperforms previous models in both speaker similarity and speech quality under the zero-shot setting. These models represent the current frontier of zero-shot TTS and serve as strong benchmarks for evaluating voice cloning robustness and defenses.



\textbf{Voice Replication Process.} As shown in the middle area of \ding{182} in Figure~\ref{fig:framework}, the voice replication process involved three key parts of the TTS model: speaker encoder $\varepsilon_{encode}\left ( \cdot  \right )$, text encoder $\varepsilon _{condition}\left ( \cdot  \right )$, and Synthesis decoder $\varepsilon_{decode}\left ( \cdot  \right )$. We denote reference voice and reference text by $\mathrm {V}_{ref}$ and $\mathrm {T}_{pt}$, respectively. We first extract the embedding of the reference voice $\mathrm{z}_{ref} = \varepsilon_{encode}\left ( \mathrm {V}_{ref};\Theta _{\varepsilon} \right )$. $\mathrm{z}_{ref}$ is used as condition to conduct alignment voice-text embedding $\mathrm{z}_{a} = \varepsilon _{condition}\left ( \mathrm {z}_{ref}, \mathrm {T}_{pt};\Theta _{ct} \right )$. $\Theta _{\varepsilon}$ and $\Theta _{ct}$ are pre-trained model parameters of speaker encoder and conditional text encoder. Then we use the decoder to synthesize the target speech: $\widehat{\mathrm {Y}} = \varepsilon_{decode}\left ( \mathrm {z}_{a};\Theta _{v} \right )$. The decoder usually consists of a vocoder and linear layers, and $\Theta _{v}$ denotes the parameters of the decoder. 

In order to present process more intuitively, we merge the conditional encoder and decoder as synthesizer $\delta_{synthesis}\left ( \cdot  \right )$. The training process can be simplified as:
\begin{equation}    \label{equ:voice cloning training}
\begin{matrix}
 \Theta_\delta = \arg \min_{\Theta } \mathbb{E}_{(\mathrm {T}_{pt}, \mathrm {V}_{ref}, \mathrm {V}_{gt}) \sim D} \mathcal{L} (\delta_{synthesis}\\    (\mathrm {T}_{pt},\varepsilon_{encode}(\mathrm {V}_{ref}); \Theta ), \mathrm {V}_{gt})
\end{matrix}
\end{equation}
where $\Theta$ is the initial parameter of the model, $\Theta_{\delta}$ is the parameter after training, and $\mathrm {V}_{gt}$ is ground truth voice. We formally define the inference process of voice cloning as: 
\begin{equation}    \label{equ:voice cloning inference}
\widehat{\mathrm {Y}} = \delta_{synthesis}\left (\varepsilon_{encode}\left ( \mathrm {V}_{ref};\Theta _{\varepsilon}\right ), \mathrm {T}_{pt}; \Theta_{\delta} \right )
\end{equation}


\subsection{Voice Privacy Protection}

Audio privacy threats arise during both training and inference stages of zero-shot voice cloning systems. During training, incorporating private data can lead to overfitting and vulnerability to backdoor attacks. This risk can be mitigated using unlearnable samples~\citep{ye2024ungeneralizable,huangunlearnable}, membership inference detection~\citep{miao2019audio,hu2022membership}, or differential privacy techniques~\citep{hassan2019differential}.

However, inference-time threats are often more immediate and pervasive, due to the accessibility and low cost of zero-shot voice cloning. Existing inference-time defenses can be grouped into three categories: audio watermarking~\citep{hua2016twenty,liu2024audiomarkbench,salah2024survey,singh2024silentcipher}, adversarial perturbations targeting ASR or speaker identification~\citep{bhanushali2024adversarial,fang2024zero,patel2025black}, and defenses against voice conversion (VC)~\citep{kang2022speaker,li2023voice,gao2025black,huang2021defending}. While each offers partial protection, none sufficiently addresses the broad threat posed by modern zero-shot voice cloning systems.
Audio Watermarking. Methods like Timbre Watermarking and AudioSeal embed imperceptible signatures into speech to support post-hoc attribution or clone detection~\citep{san2024proactive,liu2023detecting}. These approaches require prior watermark embedding and robustness to signal transformations, and become ineffective once unmarked audio is leaked or modified.
Adversarial Perturbations for ASR and Speaker ID. Other works add perturbations to obscure identity or transcription, such as VoiceBox~\citep{le2024voicebox}. However, these methods are often model-specific, fragile to architecture shifts or preprocessing, and can degrade perceptual quality, limiting practical deployment.
VC-based defenses introduce perturbations to hinder voice conversion attacks~\citep{li2023voice}. For instance, VoiceGuard uses time-domain adversarial noise guided by psychoacoustic principles to disrupt zero-shot VC. However, such defenses rely on known source-target pairs and cannot be generalized to arbitrary prompt-driven voice cloning, limiting their scope.
In summary, current methods are retrospective, brittle, or narrowly scoped. Our work fills this gap by proactively perturbing raw speech to defend against zero-shot voice cloning, without assumptions about the attacker’s model or prompt.

%% file: 3_method.tex
\section{Method}\vspace{-0.5em}
\label{method}
In this section, we first introduce the overall framework of perturbation-based defense against malicious speech cloning. In §\ref{sub:MOO}, we present a multi-objective optimization strategy that generates a universal perturbation by jointly considering a group of target audios. 
Then, we introduce a perceptual-frequency domain refinement method in §\ref{sub:Mel} that further improves the imperceptibility of the perturbation, ensuring that protected audio remains natural and indistinguishable to human listeners. 
\subsection{Perturbation-Based Voice Clone Defense Framework}
\label{sec:framework}

We introduce \textbf{CloneShield}, a perturbation-based defense framework that aims to protect a group of arbitrary audio samples from unauthorized voice cloning. The core idea is to inject imperceptible yet adversarially effective perturbations into a group of utterances, such that cloned outputs generated from these modified inputs suffer substantial degradation in speaker similarity and fidelity, while preserving naturalness for human listeners.

As illustrated in \ding{182} and \ding{183} of Figure~\ref{fig:framework}, CloneShield operates in two sequential stages. In the first stage, we generate a \emph{shared universal perturbation} by jointly optimizing over a batch of input samples from arbitrary speakers. This perturbation is crafted using a Multi-objective optimization algorithm that balances adversarial loss signals across samples~\citep{gunantara2018review}, enabling generalized protection without the need for per-sample computation, It not only saves a lot of calculations, but also improves the versatility of protection. In the second stage, we refine this universal perturbation on a per-sample basis within the perceptual-frequency domain. Specifically, we leverage multi-scale Mel-spectrogram representations to minimize audible distortions while reinforcing the degradation of cloned outputs. This process yields a final set of \emph{personalized perturbations} that combine strong defense performance with high perceptual quality. The technical details of both optimization stages are provided in §\ref{sub:MOO} and §\ref{sub:Mel}, respectively.

\subsection{Universal Perturbation via Multi-Objective Optimization}
\label{sub:MOO}
\begin{algorithm}[htbp]
\caption{Compact MGDA-based Universal Perturbation Generation}
\label{alg:mgda_uap_compact}
\begin{algorithmic}[1]
\STATE \textbf{Input:} Model $f$, inputs $\{x_i\}_{i=1}^n$, target $y_{\text{target}}$, perturbation length $L$, perturbation bound $\epsilon$, max iterations $T$, learning rate $\eta$
\STATE \textbf{Output:}Universal perturbation $\delta^*$
    \STATE $\delta \sim \mathcal{U}(-\epsilon, \epsilon)^L$, set \texttt{requires\_grad=True} {\{\color{blue} $\triangleright$ Initialize perturbation}\}
    \FOR{$t = 1$ \TO $T$}
        \STATE $\mathcal{L} = [\,\text{Loss}(f(x_i + \delta[:|x_i|]), y_{\text{target}})\,]_{i=1}^n$ {\{\color{blue} $\triangleright$ Compute losses}\}
        \STATE Solve $\min_{\bm{\alpha} \in \Delta_n} \left\| \sum_{i=1}^n \alpha_i \nabla_\delta \mathcal{L}_i \right\|_2^2$ {\{\color{blue} $\triangleright$ MGDA: For minimizing gradient conflict}\}
        \STATE $\mathcal{L}_{\text{total}} = \sum_i \alpha_i \mathcal{L}_i$ {\{\color{blue} $\triangleright$ Aggregate weighted loss}\}
        \STATE $\mathcal{L}_{\text{total}}.\text{backward()}$ 
        \STATE $\delta \leftarrow \text{clip}(\delta - \eta \cdot \delta.\text{grad}, -\epsilon, \epsilon)$ {\{\color{blue} $\triangleright$ Update and clip}\}
    \ENDFOR
\RETURN $\delta^* = \delta$
\end{algorithmic}
\end{algorithm}

In the first stage of \textbf{CloneShield}, as shown in \ding{182} of Figure \ref{fig:framework},  we aim to craft a \textit{universal perturbation} $\delta$ that protects an entire group of audio clips $\{x_1, x_2, ..., x_n\}$ from zero-shot voice cloning. Unlike instance-specific perturbations, this shared $\delta$ enables efficient deployment while maintaining defense effectiveness across diverse utterances. However, designing such a universal perturbation is challenging, as it must generalize across samples with varied speakers, durations, and content.

We formalize this task as a multi-objective optimization problem. For each sample $x_i$, we define a task-specific loss $\mathcal{L}_i$ that encourages the cloned output $f(x_i + \delta[:|x_i|])$ to deviate from a pre-defined cloning target $y_{\text{target}}$:
\begin{equation}
    \mathcal{L}_i(\delta) = \text{Loss}(f(x_i + \delta[:|x_i|]),\ y_{\text{target}}), \quad \text{for } i = 1,...,n.
\end{equation}
The optimization seeks a shared $\delta$ that minimizes all losses simultaneously:
\begin{equation}
    \min_{\delta} \left\{ \mathcal{L}_1(\delta),\ \mathcal{L}_2(\delta),\ ...,\ \mathcal{L}_n(\delta) \right\}, \quad \text{s.t.} \quad \|\delta\|_{\infty} \leq \epsilon.
\end{equation}
Solving this directly is nontrivial due to potential gradient conflicts between objectives. We adopt the \textit{Multiple Gradient Descent Algorithm} (MGDA)~\citep{desideri2012multiple} to compute a convex combination of gradients $\{\nabla \mathcal{L}_i\}$ that optimizes all tasks fairly. At each iteration, MGDA finds weights $\{\alpha_i\}$ such that the combined update direction $\sum_i \alpha_i \nabla \mathcal{L}_i$ minimizes the maximum per-task loss increase, ensuring balanced progress.

Algorithm~\ref{alg:mgda_uap_compact} outlines the procedure. We initialize a trainable perturbation vector $\delta \sim \mathcal{U}(-\epsilon, \epsilon)^L$ (Line~3), where $L$ is the perturbation length aligned with the shortest audio input. In each iteration, we apply $\delta$ to each input and compute the loss vector (Line~5), then we compute MGDA weights (Line~6). The aggregated loss $\mathcal{L}_{\text{total}}$ is backpropagated to update $\delta$ (Line~8). Then we clip the perturbation and clear the gradients before the next iteration. After $T$ steps, the resulting $\delta^*$ serves as the basis for perceptual refinement in Section~\ref{sub:Mel}.

\subsection{Mel-Spectrogram Domain Optimization for Imperceptibility}
\label{sub:Mel}

\vspace{0.5em}
\begin{algorithm}[htbp]
\caption{Mel-Spectrogram Domain Refinement with Dynamic Weighting}
\label{alg:mel_refine}
\begin{algorithmic}[1]
\STATE \textbf{Input:} TTS model $M$, benign inputs $\{x_i\}_{i=1}^n$, adversarial inputs $\{x_i^{\text{adv}}\}_{i=1}^n$, steps $T$
\STATE \textbf{Output:} Refined perturbation list $\delta^*$
\STATE $\delta \gets x_i^{\text{adv}} - x_i$, set \texttt{requires\_grad=True} {\{\color{blue} $\triangleright$ Initialize perturbation}\}
\STATE Initialize ring buffers $\texttt{ref}[5] \gets 0$, $\texttt{out}[5] \gets 0$ {\{\color{blue} $\triangleright$ Buffers for dynamic loss weighting}\}
\FOR{$i = 1$ \TO $n$}
    \FOR{$t = 1$ \TO $T$}
        \STATE $c \gets t \bmod 5$ {\{\color{blue} $\triangleright$ Circular buffer index}\}
        \STATE $x_i^{\text{adv}} \gets x_i + \delta[:|x_i|]$ {\{\color{blue} $\triangleright$ Update adversarial input}\}
        \STATE $\mathcal{L}_{\text{ref}} \gets \sum_{s \in \{512, 1024, 2048\}} \| \text{Mel}_s(x_i^{\text{adv}}) - \text{Mel}_s(x_i) \|_1$ {\{\color{blue} $\triangleright$ Multi-scale Mel similarity}\}
        \STATE $\mathcal{L}_{\text{out}} \gets \text{Dist}(M(x_i^{\text{adv}}),\ M(x_i))$ {\color{blue} $\triangleright$ Output-level divergence}
        \IF{$\texttt{ref}[c] > 0$ \textbf{and} $\texttt{out}[c] > 0$}
            \STATE $\bm{w} \gets \text{Softmax}\left([\mathcal{L}_{\text{ref}}/\texttt{ref}[c],\ \mathcal{L}_{\text{out}}/\texttt{out}[c]]\right)$ {\{\color{blue} $\triangleright$ Dynamic weight adjustment}\}
        \ELSE
            \STATE $\bm{w} \gets [0.5,\ 0.5]$ {\{\color{blue} $\triangleright$ Equal weights at initialization}\}
        \ENDIF
        \STATE $(w_{\text{ref}}, w_{\text{out}}) \gets \bm{w}$ {\{\color{blue} $\triangleright$ Split weight vector}\}
        \STATE $\mathcal{L} \gets w_{\text{ref}} \cdot \mathcal{L}_{\text{ref}} + w_{\text{out}} \cdot \mathcal{L}_{\text{out}}$ {\{\color{blue} $\triangleright$ Weighted total loss}\}
        \STATE $\texttt{ref}[c] \gets \mathcal{L}_{\text{ref}},\ \texttt{out}[c] \gets \mathcal{L}_{\text{out}}$ {\{\color{blue} $\triangleright$ Store current losses}\}
    \ENDFOR
    \STATE add $\delta$ to $\delta^*$
\ENDFOR
\RETURN $\delta^*$ {\{\color{blue} $\triangleright$ Return refined perturbation list}\}
\end{algorithmic}
\end{algorithm}

Although the universal perturbation $\delta^*$ obtained in Stage~1 successfully disrupts voice cloning, it may introduce minor artifacts that affect perceptual quality. To enhance imperceptibility without sacrificing adversarial effectiveness, we propose a second-stage refinement process that operates in the mel-spectrogram domain, which serves as a perceptual proxy aligned with human auditory sensitivity, as shown in \ding{183} of Figure \ref{fig:framework}.

The key idea is to fine-tune $\delta$ such that (i) the perturbed audio remains close to the original input in perceptual space, and (ii) the synthesized outputs remain far apart to ensure continued attack efficacy. Formally, for each input $x_i$, we define two competing objectives:
\begin{itemize}
    \item Reference loss $\mathcal{L}_\text{ref}$: the $L_1$ distance between mel-spectrograms of $x_i$ and $x_i + \delta_i$, encouraging imperceptibility;
    \item Output loss $\mathcal{L}_\text{out}$: the divergence between model outputs $M(x_i)$ and $M(x_i + \delta_i)$, encouraging defense success.
\end{itemize}

To capture both fine-grained temporal and coarse-grained spectral discrepancies, we adopt a multi-resolution mel-spectrogram decomposition using three Fourier transform scales with $n_{\text{fft}} \in \{512, 1024, 2048\}$ and the corresponding hop sizes. This allows the refinement process to consider perceptual similarity across multiple time-frequency granularities.

As outlined in Algorithm~\ref{alg:mel_refine}, stage 2 enhances the imperceptibility of adversarial perturbations through a multi-scale, perceptually aligned optimization process. Starting with the difference between each adversarial input and its benign counterpart (line 3), the algorithm iteratively updates this perturbation over $T$ steps for each sample. In each iteration (lines 4–19), it computes two key losses: a reference loss $mathcal{L}*{\text{ref}}$ (line 9), measuring multi-resolution mel-spectrogram similarity across window sizes ${512, 1024, 2048}$, and an output loss $\mathcal{L}*{\text{out}}$ (line 10), measuring the divergence between TTS model outputs. A dynamic weighting mechanism (lines 11–14) uses a circular buffer to adaptively balance these losses, enabling optimization that maintains speech quality while preserving adversarial effectiveness. The final refined perturbation for each input is collected in a list $\delta^*$ (lines 20) and returned for deployment.

%% file: 4_evaluation.tex
\section{Evaluation}\vspace{-0.4cm}
\label{evaluation}
We evaluated \textsc{CloneShield} on three representative zero-shot TTS models and five benchmark datasets, comparing against three strong baselines. The results confirm the effectiveness (\ref{sec:effectiveness}), superior performance (\ref{sec:comparison}) of our method. Additional experimental results, extended analyses, ablation study, and visualizations are provided in the appendix~\ref{sec:appendix} for completeness.

\subsection{Experiment Setup}
\label{sec: Experiment Setup}
\textbf{Model Selection}. We evaluated \textsc{CloneShield} on three state-of-the-art (SOTA) zero-shot TTS models: YourTTS~\citep{casanova2022yourtts}, XTTSv2~\citep{casanova2024xtts}, and IndexTTS~\citep{deng2025indextts}. 
YourTTS and XTTSv2 are among the most influential open-source models in the community, with the Coqui TTS framework receiving over \textbf{40K GitHub stars}, demonstrating their widespread adoption. 
YourTTS builds on VITS and supports multilingual, zero-shot voice cloning. XTTSv2 improves on this with commercial-grade zero-shot performance using only 2–6 seconds of reference audio. 
IndexTTS is the latest SOTA architecture now, achieving highly realistic zero-shot voice synthesis and setting new benchmarks in voice fidelity and speaker similarity.

\textbf{Dataset Selection}. We used five widely adopted English speech datasets in our experiments. VCTK~\citep{veaux2017vctk}, LibriSpeech ASR \citep{panayotov2015librispeech}, LibriTTS-R \citep{koizumi2023libritts}, LJSpeech~\citep{ito2017ljspeech}, and Common Voice~\citep{ardila2019commonvoice}.
These datasets cover a broad range of speaker diversity, recording conditions, and sampling rates, providing a comprehensive evaluation setting for our proposed method. 
From each dataset, we randomly select 450 utterances, ensuring diversity in speaker identity and content. These are organized into batch configurations of 1-per-batch, 3-per-batch, and 5-per-batch, with 50 batches per setting. This setup enables a systematic study of our method’s performance under different generalization pressures.
To evaluate robustness against real-world cloning scenarios, we assign distinct text prompts to each dataset. These prompts are used as target phrases during voice cloning defense scenarios, allowing us to test the perturbation effectiveness under diverse  phonetic and lexical conditions.


\textbf{Metric Selection}. To better examine the effectiveness of \textsc{CloneShield}, we introduce a comprehensive suite of evaluation metrics. We use \textit{Pyannote.audio} model \citep{bredin2020pyannote} to conduct Speaker Recognition Similarity (SRS) between original data and adversarial data. Signal-to-Distortion Ratio (SDR) \citep{yamamoto2017predicting} is applied for waveform distortion metric. We use Log Spectral Distance (LSD) \citep{swamy2020dual} and Mel Cepstral Distortion (MCD) \citep{brandt2017mel} as the metrics for estimating spectral similarity, to evaluate the destructiveness of adversarial data on timbre. The Perceptual Evaluation of Speech Quality (PESQ) \citep{martin2018deep} is employed to indicate naturalness of voice. Short-Time Objective Intelligibility (STOI) \citep{andersen2017non} and Signal-to-Noise Ratio (SNR) \citep{peng2020terahertz} are measures of perceptual-Level sample-level clarity intelligibility. We additionally apply 
\textit{SpeechRecognition 3.11.0} 
to count Character Error Rate (CER) of inputs and synthesized results. 
In addition, we also introduce the defense success rate to measure \textsc{CloneShield}'s generalizability on different speakers. We define a defense as successful if the DSR of the synthesized result falls below $0.50$. 

\subsection{Effectiveness against Zero-Shot Cloning}
\label{sec:effectiveness}




As shown in Table~\ref{tab:full_results}, \textsc{CloneShield} achieves strong protection against zero-shot voice cloning while preserving high perceptual quality on the protected inputs. Key metrics such as PESQ (up to 3.89), STOI (up to 0.977), and SDR (up to 18.75) indicate that the perturbations introduce minimal distortion, maintaining near-original naturalness and intelligibility.

In contrast, cloned outputs from these protected inputs show a substantial decline in quality: PESQ drops below 1.15, STOI falls below 0.22, and SDR becomes strongly negative (e.g., $-20.5$ for LibriTTS-R / YourTTS), confirming the effectiveness of perturbations in degrading synthesis. These trends are consistent across all evaluated models, including YourTTS, XTTSv2, and IndexTTS.
For identity leakage, the Speaker Recognition Similarity (SRS) remains above 0.9 on inputs, confirming perceptual fidelity for benign users. However, cloned outputs exhibit dramatically lower SRS scores (as low as 0.046), indicating that the synthesized voices no longer resemble the target speaker. Defense Success Rate (DSR) reaches 100\% for most cases and exceeds 92\% for XTTSv2 and IndexTTS, showing consistent and robust defense across architectures and datasets. To further evaluate the perceptual quality and the imperceptibility of our perturbations, we conducted Mean Opinion Score (MOS) tests. Detailed quantitative MOS results are provided in the Appendix~\ref{sec:appendix}. In addition to numerical ratings, qualitative feedback indicated that approximately 95\% of listeners perceived the perturbed audio as considerably close to the original audio in terms of both sound quality and timbre.

It is worth noting that these results are obtained under a challenging multi-utterance setting, where each perturbation is jointly optimized over a batch of five diverse utterances. This is more difficult than single-utterance defenses and reflects realistic deployment conditions. Additionally, due to space limits, we report three representative datasets here. Further results, including per-sample and variable-batch settings, are provided in Appendix~\ref{sec:appendix}, and confirm the generalization capability of \textsc{CloneShield} under different data and task scales.

\begin{table*}[htbp]
\centering
\caption{
Evaluation of \textsc{CloneShield} across three zero-shot TTS models and three benchmark datasets. We report performance on both the protected input audios (Input) and their corresponding cloned outputs (Output). Input metrics evaluate audio quality preservation (e.g., PESQ, STOI, SNR), while output metrics reflect cloning disruption effectiveness (e.g., increased CER, decreased speaker similarity). All results are averaged over batches of 5 utterances, representing a challenging and realistic defense scenario. CER1 is the word error rate of the original input/output, and CER2 is the word error rate of the protected input/output. A downward arrow indicates that the lower the indicator, the better the audio quality.
}

\label{tab:full_results}
\resizebox{\textwidth}{!}{
\begin{tabular}{lllrrrrrrrrrr}
\toprule
Model & Dataset & Type & CER1$\downarrow$ & CER2$\downarrow$ & MCD$\downarrow$ & PESQ$\uparrow$ & STOI$\uparrow$ & SDR$\uparrow$ & LSD$\downarrow$ & SNR$\uparrow$ & SRS & DSR \\
\midrule
\multirow{6}{*}{YourTTS~\citep{casanova2022yourtts}} 
 & \multirow{2}{*}{Common Voice~\citep{ardila2019commonvoice}} & Input &0.135 &0.157 &3.261 &3.710 &0.974 &18.755 &4.447 &9.952 &0.914 &0.000 \\ 
 & & Output &0.053 &0.839 &20.341 &1.113 &0.117 &-16.224 &6.280 &-2.790 &0.053 &1.000 \\
 & \multirow{2}{*}{LibriSpeech ASR~\citep{panayotov2015librispeech}} & Input &0.057 &0.060 &3.097 &3.899 &0.976 &18.031 &0.906 &10.070 &0.926 &0.000 \\
 & & Output &0.068 &0.853 &20.004 &1.074 &0.165 &-20.572 &8.095 &-2.297 &0.079 &1.000 \\
 & \multirow{2}{*}{LibriTTS-R~\citep{koizumi2023libritts}} & Input &0.039 &0.042 &3.344 &3.816 &0.977 &17.882 &2.343 &10.540 &0.924 &0.000 \\
 & & Output &0.012 &0.834 &21.197 &1.143 &0.122 &-20.496 &8.211 &-2.398 &0.046 &1.000 \\
\midrule
\multirow{6}{*}{XTTSv2~\citep{casanova2024xtts}} 
 & \multirow{2}{*}{Common Voice} & Input &0.131 &0.183 &2.094 &2.494 &0.962 &17.237 &4.114 &15.546 &0.878 &0.000 \\ 
 & & Output &0.161 &0.223 &19.676 &1.132 &0.199 &-17.193 &7.022 &-2.369 &0.255 &0.953 \\
 & \multirow{2}{*}{LibriSpeech ASR} & Input &0.056 &0.066 &2.432 &2.504 &0.955 &16.610 &2.375 &14.742 &0.870 &0.000 \\
 & & Output &0.148 &0.203 &19.944 &1.122 &0.224 &-17.044 &6.216 &-2.652 &0.258 &0.960 \\
 & \multirow{2}{*}{LibriTTS-R} & Input &0.038 &0.043 &2.475 &2.688 &0.967 &16.844 &2.079 &14.988 &0.903 &0.000 \\
 & & Output &0.046 &0.191 &20.416 &1.059 &0.200 &-16.984 &6.153 &-2.890 &0.237 &0.927 \\
\midrule
\multirow{6}{*}{IndexTTS} 
 & \multirow{2}{*}{Common Voice} & Input &0.138 &0.269 &2.297 &2.448 &0.952 &16.489 &4.335 &15.123 &0.815 &0.000 \\ 
 & & Output &0.007 &0.172 &19.310 &1.191 &0.169 &-17.909 &4.698 &-3.387 &0.332 &0.848 \\
 & \multirow{2}{*}{LibriSpeech ASR} & Input &0.058 &0.089 &2.530 &2.491 &0.944 &16.238 &2.317 &14.504 &0.795 &0.020 \\
 & & Output &0.001 &0.161 &14.287 &1.083 &0.248 &-15.137 &2.823 &-3.988 &0.379 &0.793 \\
 & \multirow{2}{*}{LibriTTS-R} & Input &0.039 &0.060 &2.537 &2.761 &0.959 &16.546 &1.296 &14.816 &0.837 &0.000 \\
 & & Output &0.001 &0.148 &18.138 &1.057 &0.264 &-14.187 &2.589 &-6.512 &0.288 &0.860 \\
\bottomrule
\end{tabular}
}
\end{table*}

\subsection{Comparison with Existing Defense Baselines}
\label{sec:comparison}
\vspace{-0.6em}
As shown in Table~\ref{tab:baseline}, Overall, our method achieves the best tradeoff between imperceptibility and defense robustness. On the input side, \textsc{CloneShield} maintains high audio quality and intelligibility, with PESQ, STOI, and SDR scores close to or exceeding those of the original input. Compared to watermarking approaches, which leave cloned outputs relatively unaffected, our method significantly degrades the TTS outputs. For example, on XTTSv2, our defense increases MCD to 19.88 and reduces STOI to 0.205, while AudioSeal and Timbre Watermarking remain below MCD 15 and above STOI 0.25.

On the output side, \textsc{CloneShield} consistently lowers Speaker Recognition Similarity (SRS), outperforming both adversarial and watermark-based baselines. Our method reduces SRS to as low as 0.053 and achieves 100\% defense success rate (DSR) for YourTTS, with strong performance maintained on other models. In contrast, AudioSeal and Timbre Watermarking fail to alter speaker identity features, resulting in zero DSR.

It is worth noting that when our method is compared with these three baselines, the quality of the protected input is on par with these three baselines, but the defense results are indeed far better. The visual comparison of the audio sample spectrogram is shown in Figure \ref{fig:result}. Additional breakdowns by dataset and other batch configurations are provided in Appendix~\ref{sec:appendix}.

\begin{table*}[!h]
\centering
\caption{
Comparison of \textsc{CloneShield} with three representative baselines: VoiceBox~\citep{le2024voicebox}, AudioSeal~\citep{san2024proactive}, and Timbre Watermarking~\citep{liu2023detecting}, evaluated under three TTS models. For each method, we report the impact on protected inputs (In) and the effectiveness of defense on cloned outputs (Out). CER1 and CER2 is the same meaning of above table. A downward arrow indicates that the lower the indicator, the better the audio quality. 
}

\label{tab:baseline}
\resizebox{\textwidth}{!}{
\begin{tabular}{lllrrrrrrrrrrr}
\toprule
Model & \multicolumn{1}{c}{Method} & Type & CER1$\downarrow$ & CER2$\downarrow$ & MCD$\downarrow$ & PESQ$\uparrow$ & STOI$\uparrow$ & SDR$\uparrow$ & LSD$\downarrow$ & SNR$\uparrow$ & SRS & DSR \\ 
\midrule
\multirow{8}{*}{YourTTS} 
 & \multirow{2}{*}{\centering Voicebox~\citep{le2024voicebox}} & In &0.142 &0.232 &2.357 &3.685 &0.940 &13.331 &5.200 &12.771 &0.636 &0.067 \\ 
 & & Out &0.045 &0.055 &14.068 &1.153 &0.224 &-14.727 &2.695 &-3.111 &0.539 &0.333 \\
 \cmidrule(lr){2-13}
 & \multirow{2}{*}{\centering AudioSeal~\citep{san2024proactive}} & In &0.129 &0.141 &0.514 &4.318 &0.998 &29.076 &5.255 &28.375 &0.999 &0.000 \\
 & & Out &0.045 &0.040 &10.842 &1.246 &0.395 &-9.580 &1.763 &-2.828 &0.751 &0.000 \\
 \cmidrule(lr){2-13}
 & \multirow{2}{*}{\centering Timbre Watermarking\citep{liu2023detecting}} & In &0.129 &0.155 &0.803 &3.673 &0.994 &31.242 &2.283 &30.782 &0.973 &0.000 \\
 & & Out &0.045 &0.076 &11.257 &1.187 &0.375 &-10.294 &1.914 &-2.967 &0.722 &0.000 \\
 \cmidrule(lr){2-13}
 & \multirow{2}{*}{\centering Ours} & In &0.132 &0.193 &4.061 &3.407 &0.967 &18.310 &5.226 &7.536 &0.878 &0.000 \\
 & & Out &0.045 &0.733 &18.003 &1.103 &0.130 &-17.134 &5.568 &-1.793 &0.053 &1.000 \\
\midrule
\multirow{8}{*}{XTTSv2} 
 & \multirow{2}{*}{\centering Voicebox} & In &0.142 &0.232 &2.357 &3.685 &0.940 &13.331 &5.200 &12.771 &0.636 &0.067 \\ 
 & & Out &0.063 &0.095 &13.937 &1.166 &0.258 &-13.723 &3.448 &-2.788 &0.547 &0.267 \\
 \cmidrule(lr){2-13}
 & \multirow{2}{*}{\centering AudioSeal} & In &0.129 &0.141 &0.514 &4.318 &0.998 &29.076 &5.255 &28.375 &0.999 &0.000 \\
 & & Out &0.046 &0.070 &15.444 &1.172 &0.251 &-14.959 &3.720 &-2.880 &0.752 &0.000 \\
 \cmidrule(lr){2-13}
 & \multirow{2}{*}{\centering Timbre Watermarking} & In &0.129 &0.155 &0.803 &3.673 &0.994 &31.242 &2.283 &30.782 &0.973 &0.000 \\
 & & Out &0.055 &0.110 &14.831 &1.132 &0.256 &-14.116 &3.416 &-3.083 &0.730 &0.000 \\
 \cmidrule(lr){2-13}
 & \multirow{2}{*}{\centering Ours} & In &0.136 &0.169 &2.302 &2.438 &0.960 &16.826 &4.802 &15.575 &0.868 &0.000 \\
 & & Out &0.101 &0.189 &19.884 &1.084 &0.205 &-17.717 &7.318 &-2.200 &0.202 &1.000 \\
\midrule
\multirow{8}{*}{IndexTTS} 
 & \multirow{2}{*}{\centering Voicebox} & In &0.142 &0.232 &2.357 &3.685 &0.940 &13.331 &5.200 &12.771 &0.636 &0.067 \\ 
 & & Out &0.009 &0.008 &13.177 &1.147 &0.258 &-13.086 &2.542 &-3.386 &0.592 &0.233 \\
 \cmidrule(lr){2-13}
 & \multirow{2}{*}{\centering AudioSeal} & In &0.129 &0.141 &0.514 &4.318 &0.998 &29.076 &5.255 &28.375 &0.999 &0.000 \\
 & & Out &0.009 &0.007 &12.852 &1.169 &0.296 &-13.365 &2.895 &-3.145 &0.756 &0.000 \\
 \cmidrule(lr){2-13}
 & \multirow{2}{*}{\centering Timbre Watermarking} & In &0.129 &0.155 &0.803 &3.673 &0.994 &31.242 &2.283 &30.782 &0.973 &0.000 \\
 & & Out &0.009 &0.022 &18.860 &1.397 &0.204 &-18.024 &4.087 &-2.614 &0.698 &0.033 \\
 \cmidrule(lr){2-13}
 & \multirow{2}{*}{\centering Ours} & In &0.128 &0.201 &2.443 &2.480 &0.955 &16.291 &4.931 &15.188 &0.807 &0.000 \\
 & & Out &0.011 &0.190 &20.589 &1.131 &0.160 &-17.882 &4.983 &-3.181 &0.332 &0.833 \\
\bottomrule
\end{tabular}
}
\end{table*}

\vspace{-0.4em}
\begin{figure*}[htbp]
  \centering
   \includegraphics[width=1.0\linewidth]{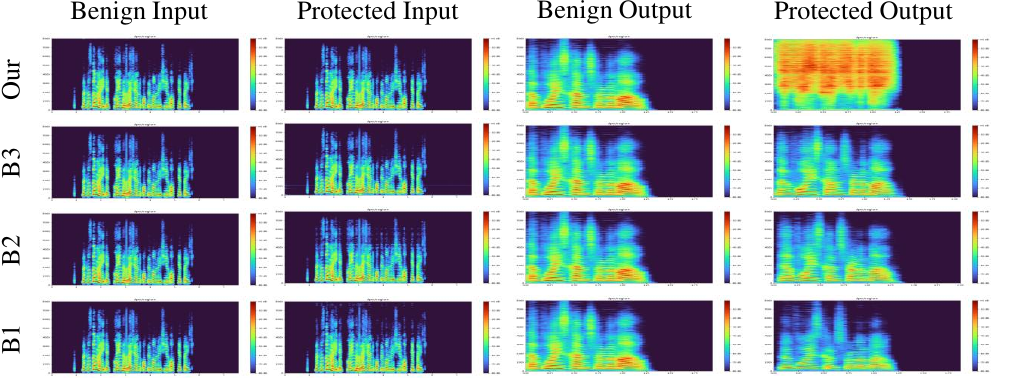}
   \caption{
We selected an audio sample for spectrogram visualization. B1, B2, and B3 represent three baseline methods: VoiceBox, AudioSeal, and Timbre Watermarking, respectively. It can be observed that our method introduces almost no perceptible difference between the protected and original audio. In contrast, the substantial discrepancy between the output of our method and the attacked output highlights the effectiveness of our defense.
}
   \label{fig:result}
\end{figure*}

%% file: 5_ethics.tex
\vspace{-0.6em}
\section{Ethics and Policy Discussion}\vspace{-0.6em}
\label{ethics}
\subsection{Ethics and Policy Needs for AI Development.} \vspace{-0.6em}
Currently, \textsc{CloneShield} achieves strong performance when tailored to individual TTS models. However, the fragmented architectures and proprietary nature of many large-scale voice synthesis systems limit the generalizability of defenses. A scalable solution requires support from TTS model providers to expose standardized defensive hooks—such as embedding-space APIs or modular encoder access—enabling unified, model-agnostic protection mechanisms. These interfaces could allow privacy-preserving perturbations to be generated without accessing internal synthesis pipelines, thereby facilitating secure, interoperable defenses without compromising model confidentiality. Realizing this vision requires policy coordination and active collaboration among AI developers, platform maintainers, and regulators.

\subsection{Lightweight Encoder-Level Perturbations for Real-World Adoption} \vspace{-0.6em}
Deploying full pipeline-level defenses across diverse TTS models can introduce significant engineering burden and concerns over system access. To balance protection and practicality, we propose an alternative \textit{encoder-only defense strategy}. Rather than intervening in the complete text-to-speech path, this strategy selectively introduces perturbations that influence the encoder’s intermediate speaker representations. It reduces integration overhead (from the full $\left \langle \Theta_{\varepsilon},\Theta_{\delta} \right \rangle$ pipeline to encoder-only $\Theta_{\varepsilon}$ access) and avoids dependency on synthesized text, making it more feasible for real-world deployments. While this may incur a slight trade-off in defense strength, it aligns better with privacy-preserving objectives and modular deployment constraints. Future work can further explore hybrid frameworks that enable scalable, cross-model, and ethically aligned voice protection. 

%% file: 6_conclusion.tex
\vspace{-0.6em} \section{Conclusion} \vspace{-0.6em}
\label{sec:conclusion}
In this paper, we present \textsc{CloneShield}, a proactive and generalizable defense framework designed to protect speaker identities against zero-shot voice cloning. Experiments on multiple state-of-the-art TTS systems and datasets demonstrate that \textsc{CloneShield} maintains naturalness and intelligibility in protected speech while significantly degrading the effectiveness of cloned outputs.
We hope this work encourages further exploration into privacy-preserving voice technologies that empower individuals to retain control over their vocal identity. The main limitation of this work is twofold: (1) testing is conducted primarily on publicly available zero-shot TTS models, and (2) evaluations are limited to offline cloning scenarios without considering real-time or few-shot attacks, both of which remain for future research.

\clearpage